\documentclass[nofootinbib]{revtex4-2}
\usepackage{amsfonts}
\usepackage{amsmath}
\usepackage{amssymb}
\usepackage{graphicx}

\begin{document}

\title{\textbf{Neutrino model with broken }$\mu -\tau $\textbf{\ Symmetry and%
} \textbf{Unflavored Leptogenesis with Dihedral Flavor Symmetry}\\
}
\author{M. Miskaoui}
\thanks{E-mail: m.miskaoui@gmail.com}

\begin{abstract}
We propose a new neutrino flavor model based on a $D_{4}\times U(1)$ flavor
symmetry providing predictions for neutrino masses and mixing along with a
successful generation of the observed Baryon Asymmetry of the Universe
(BAU). After the spontaneous breaking of the flavor symmetry, the type I
seesaw mechanism leads to a light neutrino mass matrix with broken $\mu
-\tau $ symmetry. By performing a numerical analysis, we find that the model
favors a normal mass hierarchy with the lightest neutrino mass lies in the
range $m_{1}\in \lbrack 2.516,21.351]$ \textrm{m}$\mathrm{eV}$. The
phenomenological implications of the neutrino sector are explored in detail
and the results are discussed. Moreover, the generation of BAU is addressed
via the leptogenesis mechanism from the decay of three right-handed
neutrinos $N_{i}$. Through analytical and numerical analysis of the baryon
asymmetry parameter $Y_{B}$, a successful unflavored leptogenesis takes
place within the allowed parameter space obtained from neutrino
phenomenology. We also examine interesting correlations between $Y_{B}$ and
low energy observables and provide a comprehensive discussion of the results.

\end{abstract}

\flushbottom
\affiliation{Faculty of \ Science,
Mohammed V University, Rabat, Morocco} 
\maketitle
\section{Introduction}

Despite its unrivaled success in describing fundamental interactions, the
standard model (SM) falls short of providing a comprehensive explanation for
phenomena related to neutrinos. These include the origin of neutrino masses,
lepton flavor mixing, and the matter-antimatter asymmetry observed in the
Universe. The discovery of neutrino oscillations is one of the most
significant experimental discoveries in recent years, confirming a nonzero
neutrino masses and thus providing clear evidence for physics beyond the
standard model (BSM) \cite{A1,A2,E1,E2,E3}. In light of this, several
scenarios have been proposed to explain the origin of the nonzero neutrino
masses, with the simplest of these involves extending the SM by adding three
right-handed neutrinos (RH). This extension incorporates the type I seesaw
mechanism, which generates small neutrino masses \cite{A3,A4,A5,A6,A7}.
Nevertheless, this mechanism faces limitations in its ability to reproduce
the observed values of the oscillation parameters, namely the mixing angle $%
\theta _{ij}$, the mass-squared differences $\Delta m_{ij}^{2}$ and $CP$
phase $\delta _{CP}$. A summary of the latest status of different neutrino
oscillation parameters is summarized in Table (\ref{t0}) \cite{A8}.
\begin{table}[h]
\centering$%
\begin{tabular}{l|l|l|}
\cline{2-3}\cline{2-3}
& \ \ $\ \ \left.
\begin{array}{c}
\text{Normal Hierarchy} \\
\text{Best fit }(-3\sigma \rightarrow +3\sigma )%
\end{array}%
\right. $\  & \ \ \ $\left.
\begin{array}{c}
\text{Inverted Hierarchy} \\
\text{Best fit }(-3\sigma \rightarrow +3\sigma )%
\end{array}%
\right. $ \\ \hline\hline
\multicolumn{1}{|l|}{$\sin ^{2}\theta _{13}$} & $0.02219(0.02032\rightarrow
0.02410)$ & $0.02238(0.02052\rightarrow 0.02428)$ \\ \hline
\multicolumn{1}{|l|}{$\sin ^{2}\theta _{12}$} & $\ \ \ \ \
0.304(0.269\rightarrow 0.343)$ & $\ \ \ \ \ \ 0.304(0.269\rightarrow 0.343)$
\\ \hline
\multicolumn{1}{|l|}{$\sin ^{2}\theta _{23}$} & $\ \ \ \ \
0.573(0.415\rightarrow 0.616)$ & $\ \ \ \ \ \ 0.575(0.419\rightarrow 0.617)$
\\ \hline
\multicolumn{1}{|l|}{$\Delta m_{21}^{2}/10^{-5}$} & $\ \ \ \ \ \ \ \
7.42(6.82\rightarrow 8.04)$ & $\ \ \ \ \ \ \ \ \ 7.42(6.82\rightarrow 8.04)$
\\ \hline
\multicolumn{1}{|l|}{$\Delta m_{3l}^{2}/10^{-3}$} & $\ \ \ \ \ \
2.517(2.435\rightarrow 2.598)$ & $\ -2.498(-2.581\rightarrow -2.414)$ \\
\hline
\multicolumn{1}{|l|}{$\delta _{CP}^{\circ }$} & $\ \ \ \ \ \ \ \ \ \ \
197(120\rightarrow 369)$ & $\ \ \ \ \ \ \ \ \ \ \ \ \ 282(193\rightarrow
352) $ \\ \hline
\end{tabular}%
$%
\caption{Best-fit values and $3\protect\sigma $ allowed ranges of the
neutrino oscillation parameters where $l=1$ for Normal Hierarchy (NH) and $%
l=2$ for Inverted Hierarchy (IH); taken from Ref. \protect\cite{A8}.}
\label{t0}
\end{table}
Over the years, a range of mixing patterns have been proposed in response to
advances in neutrino oscillation data. One such pattern is the Trimaximal
mixing (TM$_{\mathrm{2}}$), which is regarded as a powerful scheme for
describing neutrino mixing due to its ability to produce mixing angles that
are in agreement with current oscillation data \cite%
{A9,A10,A11,A12,A13,A14,A15}. Therefore, it is compelling to look for a
common framework that can explain both the smallness of neutrino masses and
the large mixing angles. From this perspective, non Abelian discrete
symmetries offer a new promising approach for understanding the flavor
structure of leptons and quarks. These new symmetries play a crucial role in
establishing connections among different fermion generations. In particular,
when considering discrete groups with triplet representations, it is assumed
that the three generations of fermions transform as a triplet under these
symmetries. Remarkably, discrete groups such as $A_{4}$, which exhibit this
characteristic, has been extensively used in the literature to provide a
theoretical origin to the neutrino masses and mixing. In the very early
attempts \cite{A16,A17,A18,A19}, it was employed in type I seesaw models to
accommodate the tribimaximal mixing (TBM) pattern \cite{A20,A21}. However,
these models fail to fit the latest observations concerning the reactor
angle $\theta _{13}$ \cite{A22,A23,A24}, triggering subsequently
modifications to the original approach by one of the following strategies;
\emph{(i)} introducing a small perturbations to the tribimaximal mixing
angles using gauge singlet scalar fields called flavons \cite{A25,A26,Y0} or
\emph{(ii)} considering deviations from the TBM pattern by incorporating
corrections in the charged leptons \cite{A27}.\newline
To successfully account for the observed oscillation data, an alternative
approach can be pursued by constructing flavor models using discrete groups
with doublet representations, such as $S_{3}$ and $D_{4}$. In this
scenarios, the three generations of neutrinos would be associated with
singlet and doublet representations, rather than a triplet. The $D_{4}$\
discrete group in particular has been used recently in various frameworks.
For example, it has been implemented as a flavor symmetry in the $SU(5)$ GUT
to address fermion masses and mixing\footnote{%
A systematic study of the dihedral group $D_{n}$ as flavor symmetry to
understand the lepton and quark mixing patterns have been performed in Refs.
\cite{X1,X2}.} \cite{A28,A29,X3}. Additionally, it has been employed in
orbifold models derived from heterotic strings \cite{A30,A31,A32}, as well
as in building viable Minimal Supersymmetric Standard Model (MSSM)-like
prototypes in F-theory \cite{A33,A34,A35}. In the SM framework, the $D_{4}$
flavor model was initially proposed by Grimus and Lavoura to accommodate the
observed neutrino masses and mixing \cite{A36,A37,A38,E4}. Their study
demonstrated that the $D_{4}$ flavor symmetry naturally predicts the $\mu
-\tau $ symmetry in the neutrino mass matrix, leading to a maximal
atmospheric angle $\theta _{23}=\frac{\pi }{4}$ and a vanishing reactor
angle $\theta _{13}=0$, while the solar angle $\theta _{21}$\ remains
arbitrary, but generally expected to be large\footnote{%
The fact that the value of the reactor angle $\theta _{13}$ is small, the $%
\mu -\tau $ symmetry in the neutrino mass matrix can still be considered
valid at leading order.}. A supersymmetric (SUSY) versions of $D_{4}$\ model
has been proposed in Refs. \cite{A39,A40,A41,E5}, where irrespective to the
mechanism employed to generate tiny mass of neutrinos, they lead to similar
predictions regarding the mixing angles with $\theta _{23}=\frac{\pi }{4}$
and $\theta _{13}=0$. It is worth mentioning that the studies in Refs. \cite%
{A40,A41} have concisely explored small deviations from $\theta _{13}=0$ and
$\theta _{23}=\frac{\pi }{4}$, however, no thorough study concerning the
neutrino phenomenology has been performed. \newline
The $D_{4}$ flavor symmetry group has been utilized in various extensions of
the SM with the aim of providing viable predictions for fermion mass and
mixing hierarchiy. This symmetry group has been employed to address the
hierarchy of fermion masses and mixing angles within models featuring two
\cite{E6}, three \cite{E7}, and four Higgs doublets \cite{E8}. Moreover, it
has been investigated within the $B-L$ extension of the SM to elucidate mass
spectra and mixing parameters concerning charged leptons and/or quarks \cite%
{E9,E10}. Additionally, the incorporation of $D_{4}$ flavor symmetry has
been explored within the 3-3-1 model \cite{E11,E12,E13,E14}, as well as in
neutrino models, to stabilize dark matter and generate small neutrino masses
\cite{E15,A42}\footnote{%
Radiative neutrino mass model with $D_{4}$ discrete group has been studied
in Ref. \cite{A42}.}.

Despite significant evidence suggesting a matter-antimatter imbalance in the
universe, its source remains a mystery. The problem of explaining the origin
of this matter-antimatter asymmetry is known as the baryogenesis problem.
The type I seesaw mechanism, in addition to being responsible for the
generation of tiny neutrino masses, also provides an explanation for the
observed matter-antimatter asymmetry of the universe via the leptogenesis
mechanism \cite{A43}. In this scenario, the lepton asymmetry generated by
the out-of-equilibrium decays of the RH neutrinos is converted into a baryon
asymmetry by sphaleron processes and explains eventually the
matter-antimatter asymmetry in the universe \cite{A44}. Implementing
leptogenesis in the seesaw models with $D_{4}$ discrete symmetry has been
explored within the framework of supersymmetric $SU(5)$ grand unified models
\cite{A29,X3}. However, further investigations into its application within
the MSSM framework is still to be explored.

In this work, we propose a new neutrino model based on the dihedral discrete
group $D_{4}$ to address the problems of neutrino masses, mixing and the
generation of the BAU. To the best of our knowledge, this is the first $%
D_{4} $ flavor model that has been proposed in the MSSM framework. It has
the ability to provide a simultaneous explanation for all these issues while
maintaining consistency with the existing observational and experimental
data.To address the charged lepton mass hierarchy and differentiate it from
the neutrino sector, the $D_{4}$ discrete group is supplemented by an extra
global $U(1)$ symmetry. In the charged lepton sector, the resulting mass
matrix is diagonal, and thus the leptonic mixing arises from the neutrino
sector. Within the neutrino sector, the type I seesaw mechanism is employed
to achieve small neutrino masses, while the TM$_{\mathrm{2}}$ mixing pattern
describes the neutrino mixing after the flavor symmetry gets broken when
flavon acquire VEVs. This sector involves five free parameters that have to
be fixed in order to provide predictions on the observables $\theta _{ij}$, $%
\Delta m_{ij}$\ and $\delta _{CP}$ within their $3\sigma $ experimental
ranges. Moreover, predictions on the absolute neutrino mass scale are
extrapolated by investigating the non oscillation observables namely; the
electron neutrino mass $m_{\beta }$ from beta decay experiments, the
effective Majorana mass $m_{\beta \beta }$ from neutrinoless double beta
decay ($0\nu \beta \beta $) experiments and the sum of the three active
neutrino masses $\Sigma m_{i}$ from cosmological observations. Leptogenesis,
on the other hand, cannot be generated if only leading order contributions
to the Dirac Yukawa matrix are considered. Therefore, a higher order
correction involving a new scalar flavon field is taken into account. In
this regard, we estimate the baryon asymmetry $Y_{B}$ in the unflavored
approximation from the decays of three RH neutrinos $N_{1,2,3}$. The model
yields the following main predictions:

\begin{itemize}
\item the normal hierarchy for the neutrino mass spectrum is preferred,

\item the reactor angle has a nonzero value $\theta _{13}\neq 0$,

\item the atmospheric angle lies in the lower octant $\theta _{23}<\frac{\pi
}{4}$,

\item the obtained values of $m_{\beta \beta }$ are testable at future $0\nu
\beta \beta $ experiments,

\item the baryon asymmetry parameter $Y_{B_{1}}$, arising from the decay of
the RH neutrino $N_{1}$, emerges as the primary contribution to produce the
observed BAU and

\item the high energy phase $\phi _{\omega }$\ ---originates from the extra
contribution in the Dirac Yukawa matrix--- provides a new source of $CP$\
violation.
\end{itemize}

The layout of the article is as follows. In section \ref{sec2}, we provide
the necessary components for building the $D_{4}\times U(1)$ model and
deriving the mass matrices of the charged lepton and neutrino sectors. In
addition, we numerically investigate the parameter space of the model to
satisfy the recent $3\sigma $ regions of the neutrino oscillation
parameters. In section \ref{sec3}, we thoroughly examine the prediction of
the absolute neutrino mass scale from non oscillatory experiments. In
section \ref{sec4}, we investigate leptogenesis in the current setup to
explain the baryon asymmetry of the universe. Finally, we give the
conclusions in section \ref{sec5}. The paper includes three appendices:
Appendix \ref{app1} provides some algebraic tools on the\ dihedral group $%
D_{4}$. In Appendix \ref{app2} we study the minimization of the scalar
superpotential, which eventually leads to the desired alignments of the
flavon doublet VEVs. In Appendix \ref{app3} we briefly explore the effect of
the NLO correction to the Dirac Yukawa matrix operator on the neutrino
masses and mixing.

\section{Constraining parameters from neutrino oscillation data}

\label{sec2}

\subsection{Structure of the model}

In this study, we explore an extension of the MSSM that incorporates the $%
D_{4}\times U(1)$ flavor symmetry and three RH neutrinos to generate lepton
masses and mixing. The auxiliary $U(1)$ symmetry is introduced to separate
the neutrino from the charged lepton sectors as well as to achieve the
desired mass matrices. For this purpose, we consider that the three RH
neutrinos $N_{i=1,2,3}^{c}$ and the three left-handed leptons $L_{e,\mu
,\tau }$ transform under $D_{4}$ as $\mathbf{1}_{+,+}\oplus \mathbf{2}_{0,0}$
with the same $U(1)$ quantum numbers. Along with that, the right-handed
charged leptons $l_{e}^{c},$ $l_{\mu }^{c}$\ and $l_{\tau }^{c}$ transform
as singlets under the discrete symmetry $D_{4}$, while having different $%
U(1) $ charges. This is essential in order to induce the mass hierarchy
among the charged leptons and so that the lepton mixing in our model results
mainly from the neutrino sector. As for the scalar sector, the usual MSSM
Higgs doublets $H_{u}$\ and\ $H_{d}$ transform trivially under $D_{4}$, but
have different $U(1)$ charges. The transformation properties of matter and
Higgs fields under the flavor symmetry $D_{4}\times U(1)$\ are depicted in
Table (\ref{t2}).
\begin{table}[h]
\centering$%
\begin{tabular}{|l||l|l|l|l|l|l|l|l|l|}
\hline\hline
Fields & $L_{e}$ & $(L_{\mu },L_{\tau })$ & $l_{e}^{c}$ & $l_{\mu }^{c}$ & $%
l_{\tau }^{c}$ & $N_{1}^{c}$ & $N_{3,2}^{c}$ & $H_{u}$ & $H_{d}$ \\
\hline\hline
$SU(3)_{C}\times SU(2)_{L}\times U(1)_{Y}$ & $(1,2)_{-\frac{1}{2}}$ & $%
(1,2)_{_{-\frac{1}{2}}}$ & $(1,1)_{1}$ & $(1,1)_{1}$ & $(1,1)_{1}$ & $%
(1,1)_{0}$ & $(1,1)_{0}$ & $(1,2)_{_{\frac{1}{2}}}$ & $(1,2)_{\frac{1}{2}}$
\\ \hline
$D_{4}$ & $1_{+,+}$ & $\ \ \ 2_{0,0}$ & $1_{+,+}$ & $1_{+,+}$ & $1_{+,+}$ & $%
1_{+,+}$ & $2_{0,0}$ & $1_{+,+}$ & $1_{+,+}$ \\ \hline
$U(1)$ & $-2$ & \ $\ \ -2$ & $-2$ & $5$ & $-4$ & $-1$ & $-1$ & $3$ & $1$ \\
\hline\hline
\end{tabular}%
$%
\caption{Transformation properties of matter and Higgs fields under $%
SU(3)_{C}\times SU(2)_{L}\times U(1)_{Y}$ gauge symmetry and $D_{4}\times
U(1)$ flavor symmetry. The two generations of RH neutrinos are hosted by the
$D_{4}$ doublet as $N_{3,2}^{c}=(N_{3}^{c},N_{2}^{c})^{T}$.}
\label{t2}
\end{table}

We introduce in our model eight flavon superfields in order to break the $%
D_{4}\times U(1)$ flavor symmetry and to engineer the invariance of the
superpotentials in the charged lepton and neutrino sectors. For the charged
lepton sector, we introduce three flavons $\phi $, $\chi $\ and $\psi $
which transform differently under the flavor symmetry. Their assignments are
chosen in such a way to prevent their coupling in the neutrino sector and to
induce the hierarchical structure of the three generations of charged
leptons. As for the neutrino sector, we incorporate five flavons denoted as $%
\rho _{1}$, $\rho _{2}$, $\rho _{3}$, $\eta $ and $\sigma $ in order to
accommodates the observed neutrino oscillation data. These flavon fields
break the flavor symmetry once they acquire VEVs along suitable directions.
Table (\ref{t3}) summarizes their respective quantum numbers under $%
D_{4}\times U(1)$
\begin{table}[h]
\centering$%
\begin{tabular}{|l||l|l|l|l|l||l|l|l|}
\hline\hline
Flavons & $\rho _{1}$ & $\rho _{2}$ & $\rho _{3}$ & $\eta $ & $\sigma $ & $%
\phi $ & $\chi $ & $\psi $ \\ \hline\hline
$SU(3)_{C}\times SU(2)_{L}\times U(1)_{Y}$ & $(1,1)_{0}$ & $(1,1)_{0}$ & $%
(1,1)_{0}$ & $(1,1)_{0}$ & $(1,1)_{0}$ & $(1,1)_{0}$ & $(1,1)_{0}$ & $%
(1,1)_{0}$ \\ \hline
$D_{4}$ & $1_{+,+}$ & $1_{+,-}$ & $1_{-,+}$ & $2_{0,0}$ & $2_{0,0}$ & $%
1_{+,+}$ & $2_{0,0}$ & $2_{0,0}$ \\ \hline
$U(1)$ & $2$ & $2$ & $2$ & $2$ & $2$ & $3$ & $-4$ & $5$ \\ \hline\hline
\end{tabular}%
$%
\caption{Transformation properties of the flavon superfields under $%
SU(3)_{C}\times SU(2)_{L}\times U(1)_{Y}$ gauge symmetry and $D_{4}\times
U(1)$ flavor symmetry.}
\label{t3}
\end{table}

\begin{itemize}
\item \textbf{Charged lepton sector}
\end{itemize}

With the above mentioned fields and their respective charge configurations,
the superpotential relevant for the charged lepton masses at leading order
(LO) reads as%
\begin{equation}
\mathcal{W}_{l}=\lambda _{e}\frac{\phi }{\Lambda }L_{e}l_{e}^{c}H_{d}+%
\lambda _{\mu }\frac{\chi }{\Lambda }L_{\mu ,\tau }l_{\mu }^{c}H_{d}+\lambda
_{\tau }\frac{\psi }{\Lambda }L_{\mu ,\tau }l_{\tau }^{c}H_{d}  \label{Wl}
\end{equation}%
where $\lambda _{e}$, $\lambda _{\mu }$ and $\lambda _{\tau }$ are the
coupling constants associated with the three charged leptons and $\Lambda $\
is the flavor symmetry breaking scale. Furthermore, when the Higgs field $%
H_{d}$ and the flavon fields acquire their VEVs in the following directions%
\footnote{%
Recall that the vacuum expectation values $\upsilon _{u}$ and $\upsilon _{d}$
of the usual Higgs superfields, $H_{u}$ and $H_{d}$, are related to the SM
Higgs VEV as $\upsilon _{H}^{2}=\upsilon _{u}^{2}+$ $\upsilon _{d}^{2}$
while the ratio of Higgs superfields is $\tan \beta $ $=\upsilon
_{u}/\upsilon _{d}$.}%
\begin{equation}
\left\langle H_{d}\right\rangle =\left(
\begin{array}{c}
0 \\
\upsilon _{d}%
\end{array}%
\right) \quad ,\quad \left\langle \phi \right\rangle =\upsilon _{\phi }\quad
,\quad \left\langle \chi \right\rangle =(0,\upsilon _{\chi })^{T}\quad
,\quad \left\langle \psi \right\rangle =(\upsilon _{\psi },0)^{T}
\label{vev1}
\end{equation}%
the charged lepton Yukawa matrix takes the diagonal form as%
\begin{equation}
\mathcal{Y}_{L}=\left(
\begin{array}{ccc}
\lambda _{e}\frac{\upsilon _{\phi }}{\Lambda } & 0 & 0 \\
0 & \lambda _{\mu }\frac{\upsilon _{\chi }}{\Lambda } & 0 \\
0 & 0 & \lambda _{\tau }\frac{\upsilon _{\psi }}{\Lambda }%
\end{array}%
\right)  \label{yl}
\end{equation}%
As result, the masses of the three charged leptons are given by%
\begin{equation}
m_{e}=\lambda _{e}\frac{\upsilon _{d}\upsilon _{\phi }}{\Lambda }\quad \text{%
,}\quad m_{\mu }=\lambda _{\mu }\frac{\upsilon _{d}\upsilon _{\chi }}{%
\Lambda }\quad \text{,}\quad m_{\tau }=\lambda _{\tau }\frac{\upsilon
_{d}\upsilon _{\psi }}{\Lambda }
\end{equation}%
It is clear that the charged lepton mass hierarchies are generated after the
spontaneous flavor symmetry breaking. Indeed, assuming that the coupling
constants $\lambda _{e}$, $\lambda _{\mu }$, and $\lambda _{\tau }$ have the
same order of magnitude, the mass hierarchy of the charged leptons can be
expressed in terms of the Wolfenstein parameter $\lambda $ as \cite{X4}
\begin{equation}
m_{e}:m_{\mu }:m_{\tau }\sim \lambda ^{4}:\lambda ^{2}:1
\end{equation}%
This is ensured by the hierarchy of the VEVs of the flavon fields, given by:
$\upsilon _{\phi }:\upsilon _{\chi }:\upsilon _{\psi }\sim \lambda
^{4}:\lambda ^{2}:1$. On the other hand, since the charged lepton mass
matrix in Eq. (\ref{yl}) is diagonal, the three mixing angles in the charged
lepton sector vanish $\theta _{ij}^{l}=0$. Therefore, the leptonic mixing
results mainly from the neutrino sector as we will see in subsequent
paragraph.

\begin{itemize}
\item \textbf{Neutrino sector}
\end{itemize}

The light neutrino masses are generated through the type I seesaw mechanism
given by the formula $m_{\nu }=m_{D}m_{M}^{-1}m_{D}^{T}$. With respect to
the invariance under $D_{4}\times U(1)$\ flavor symmetry, the relevant
superpotential for neutrino mass generation is given by%
\begin{eqnarray}
\mathcal{W}_{\nu } &=&\lambda _{1}N_{1}^{c}L_{e}H_{u}+\lambda
_{2}N_{3,2}^{c}L_{\mu ,\tau }H_{u}+\lambda _{3}N_{1}^{c}N_{1}^{c}\rho
_{1}+\lambda _{4}N_{3,2}^{c}N_{3,2}^{c}\rho _{1}  \notag \\
&&+\lambda _{5}N_{1}^{c}N_{3,2}^{c}\eta +\lambda
_{6}N_{1}^{c}N_{3,2}^{c}\sigma +\lambda _{7}N_{3,2}^{c}N_{3,2}^{c}\rho
_{2}+\lambda _{8}N_{3,2}^{c}N_{3,2}^{c}\rho _{3}  \label{Wn}
\end{eqnarray}%
where $\lambda _{i=1,...,8}$ are Yukawa coupling constants. The first two
terms are the Dirac Yukawa couplings leading to the Dirac mass matrix $m_{D}$%
\ while the remaining couplings give rise to the Majorana mass matrix $m_{M}$%
. Similarly, the Higgs doublet develops its VEV as usual $\left\langle
H_{u}\right\rangle =\left(
\begin{array}{c}
\upsilon _{u} \\
0%
\end{array}%
\right) $, while the VEV alignments of the flavons are chosen as follows
\begin{eqnarray}
\left\langle \rho _{1}\right\rangle &=&\upsilon _{\rho _{1}}\quad ,\quad
\left\langle \rho _{2}\right\rangle =\upsilon _{\rho _{2}}\quad ,\quad
\left\langle \rho _{3}\right\rangle =\upsilon _{\rho _{3}} \\
\left\langle \eta \right\rangle &=&(\upsilon _{\eta },\upsilon _{\eta
})^{T}\quad ,\quad \left\langle \sigma \right\rangle =(0,\upsilon _{\sigma
})^{T}.  \label{vev2}
\end{eqnarray}%
Using the tensor product of $D_{4}$\ irreducible representations given in
Eqs. (\ref{c1}) and (\ref{c2}), this leads to the Dirac $m_{D}$ and
Majorana\ $m_{M}$ neutrino mass matrices at the LO as follows%
\begin{equation}
m_{D}=\upsilon _{u}\left(
\begin{array}{ccc}
\lambda _{1} & 0 & 0 \\
0 & \lambda _{2} & 0 \\
0 & 0 & \lambda _{2}%
\end{array}%
\right) \quad ,\quad m_{M}=\left(
\begin{array}{ccc}
\lambda _{3}\rho _{1} & \lambda _{5}\eta & \lambda _{5}\eta +\lambda
_{6}\sigma \\
\lambda _{5}\eta & \lambda _{7}\rho _{2}-\lambda _{8}\rho _{3} & 2\lambda
_{4}\rho _{1} \\
\lambda _{5}\eta +\lambda _{6}\sigma & 2\lambda _{4}\rho _{1} & \lambda
_{7}\rho _{2}+\lambda _{8}\rho _{3}%
\end{array}%
\right)  \label{DM}
\end{equation}%
Since the Dirac matrix $m_{D}$ is diagonal, the neutrino mixing arises
mainly from the Majorana mass matrix $m_{M}$. Moreover, assuming that the
coupling constants $\lambda _{1}$ and $\lambda _{2}$ are of the same order,
then the mass matrices $m_{M}$\ and $m_{\nu }=m_{D}m_{M}^{-1}m_{D}^{T}$ are
identical from the point of view of symmetry. As a result, the neutrino
mixing matrix $\mathcal{U}_{\nu }$ can be obtained from the diagonalization
of Majorana mass matrix as $m_{M}^{diag}=\mathcal{U}_{\nu }^{\dagger }m_{M}%
\mathcal{U}_{\nu }$ where $m_{M}$ can be expressed as the sum of two
matrices as follows%
\begin{equation}
m_{M}=m_{M_{1}}+m_{M_{2}}=\left(
\begin{array}{ccc}
\lambda _{3}\rho _{1} & \lambda _{5}\eta & \lambda _{5}\eta \\
\lambda _{5}\eta & 0 & 2\lambda _{4}\rho _{1} \\
\lambda _{5}\eta & 2\lambda _{4}\rho _{1} & 0%
\end{array}%
\right) +\left(
\begin{array}{ccc}
0 & 0 & \lambda _{6}\sigma \\
0 & \lambda _{7}\rho _{2}-\lambda _{8}\rho _{3} & 0 \\
\lambda _{6}\sigma & 0 & \lambda _{7}\rho _{2}+\lambda _{8}\rho _{3}%
\end{array}%
\right)  \label{Majm1}
\end{equation}%
The leptonic mixing primarily arises from the neutrino sector due to the
diagonal structure of the charged lepton mass matrix in Eq. (\ref{yl}).
Moreover, to build a viable neutrino mass model, it is crucial for the
neutrino mass matrix to exhibit a broken $\mu -\tau $ symmetry. In this
respect, the matrix $m_{M_{1}}$, which involves the flavons $\rho _{1}$ and $%
\eta $, naturally exhibits the $\mu -\tau $ symmetry, leading to a maximal
atmospheric angle $\theta _{23}=\frac{\pi }{4}$ and a vanishing reactor
angle $\theta _{13}=0$ \cite{B1,B2,B3,B4,B5}. Furthermore, the matrix $%
m_{M_{2}}$, which breaks the $\mu -\tau $ symmetry in the Majorana matrix $%
m_{M}$ through the flavon fields $\rho _{2,3}$ and $\sigma $, induces a
nonzero $\theta _{13}$ and a deviation of $\theta _{23}$ from its maximal
value. These features ensure the viability of the neutrino sector within our
model.The determination of the deviation from the values $\theta _{13}=0$\
and $\theta _{23}=\frac{\pi }{4}$ is not well-defined when considering a
straightforward analysis in the neutrino sector without specific mixing
patterns. However, by investigating the neutrino mass matrix within specific
mixing patterns, such as trimaximal mixing, a more accurate determination of
the deviation can be achieved. This is because such patterns is consistent
with the broken $\mu -\tau $ symmetry in the neutrino mass matrix and
establish connections between the mixing angles $\theta _{ij}$ and deviation
parameter $\theta $, as we will investigate in the following analysis.%
\newline
Before we study the model predictions regarding neutrino masses and mixing,
let us recall some properties of TBM mixing pattern and its deviation
leading to TM$_{2}$. When the neutrino mass matrix $m_{\nu }$ satisfies both
the $\mu -\tau $ symmetry and the condition $(m_{\nu })_{11}+(m_{\nu
})_{12}-(m_{\nu })_{22}=(m_{\nu })_{23}$ among its entries, the resulting $%
m_{\nu }$ mass matrix can be diagonalized by TBM. The introduction of a
small correction $\delta m_{\nu }$ to the neutrino mass matrix $m_{\nu }$
allows for a deviation from TBM. When this matrix perturbation $\delta
m_{\nu }$ breaks the original $\mu -\tau $ symmetry while simultaneously
inducing the magic symmetry in the resulting neutrino mass matrix,\ $m_{\nu
} $\ becomes consistent with trimaximal mixing (TM)\footnote{%
The magic symmetry refers to the property where the sums of the rows and
columns of the neutrino mass matrix are equal \cite{B0}.}. Two specific
matrix perturbations, denoted as $\delta m_{\nu }^{1}$ and $\delta m_{\nu
}^{2}$, have been identified as leading to Trimaximal Mixing (TM$_{2}$),
they are defined as follows%
\begin{equation}
\delta m_{\nu }^{1}=\left(
\begin{array}{ccc}
0 & 0 & \mathrm{k} \\
0 & \mathrm{k} & 0 \\
\mathrm{k} & 0 & 0%
\end{array}%
\right) \quad ,\quad \delta m_{\nu }^{2}=\left(
\begin{array}{ccc}
0 & \mathrm{k} & 0 \\
\mathrm{k} & 0 & 0 \\
0 & 0 & \mathrm{k}%
\end{array}%
\right)  \label{per}
\end{equation}%
where $\mathrm{k}$\ is\ a deviation parameter which is expected to be small;
it is responsible for inducing a nonzero reactor angle $\theta _{13}\neq 0$
and non maximal atmospheric angle $\theta _{23}\neq 45^{\circ }$. Returning
to our model, the magic symmetry is manifested in the Majorana matrix $m_{M}$
(see Eq.(\ref{Majm1})) by reducing the number of free parameters through the
assumptions $2\lambda _{4}\rho _{1}=\lambda _{3}\rho _{1}+\lambda _{5}\eta $
\ and $\lambda _{7}\upsilon _{\rho _{2}}=-\lambda _{8}\upsilon _{\rho
_{3}}=\lambda _{6}\upsilon _{\sigma }/2$. As a result, the Dirac and
Majorana mass matrices in Eq. (\ref{DM}) reduce to%
\begin{equation}
m_{D}=\upsilon _{u}\left(
\begin{array}{ccc}
\lambda _{1} & 0 & 0 \\
0 & \lambda _{1} & 0 \\
0 & 0 & \lambda _{1}%
\end{array}%
\right) \quad ,\quad m_{M}=\Lambda \left(
\begin{array}{ccc}
a & b & b \\
b & 0 & a+b \\
b & a+b & 0%
\end{array}%
\right) +\Lambda \left(
\begin{array}{ccc}
0 & 0 & \mathrm{k} \\
0 & \mathrm{k} & 0 \\
\mathrm{k} & 0 & 0%
\end{array}%
\right)
\end{equation}%
where for clarity, we introduce the notations $a=\frac{\lambda _{3}\upsilon
_{\rho _{1}}}{\Lambda }$, $b=\frac{\lambda _{5}\upsilon _{\eta }}{\Lambda }$
and the deviation parameter $\mathrm{k}$ is given in terms of the flavon VEV
$\upsilon _{\sigma }$ as $\mathrm{k}=\frac{\lambda _{6}\upsilon _{\sigma }}{%
\Lambda }$. The total neutrino mass matrix $m_{\nu }$ is now calculated
using the type I seesaw mechanism $m_{\nu }=m_{D}m_{M}^{-1}m_{D}^{T}$ as
follows%
\begin{equation}
m_{\nu }=\frac{m_{0}}{H}\left(
\begin{array}{ccc}
-\left( a+b\right) ^{2} & \left( a+b\right) \left( b+\mathrm{k}\right) &
b^{2}-\mathrm{k}^{2}-b\left( \mathrm{k}-a\right) \\
\left( a+b\right) \left( b+\mathrm{k}\right) & -\left( b+\mathrm{k}\right)
^{2} & -a^{2}-ab+b^{2}+\mathrm{k}b \\
b^{2}-\mathrm{k}^{2}-b\left( \mathrm{k}-a\right) & -a^{2}-ab+b^{2}+\mathrm{k}%
b & a\mathrm{k}-b^{2}%
\end{array}%
\right)  \label{mn}
\end{equation}%
with $m_{0}=\frac{\left( \lambda _{1}\upsilon _{u}\right) ^{2}}{\Lambda }$
and $H=\left( a+2b+\mathrm{k}\right) \left( a\mathrm{k}-a^{2}+b^{2}-\mathrm{k%
}^{2}\right) $. In our setup, the presence of the parameter\textrm{\ k }
---which is derived from additional terms in the superpotential (\ref{Wn}%
)--- ensures $\mu -\tau $\textrm{\ }symmetry breaking in the neutrino mass
matrix, and subsequently a small deviation from the TBM values of the mixing
angles. Besides,\textrm{\ }to ensure $CP$ violation in the lepton sector,
the complex nature of the parameters $a$, $b$ and \textrm{k} has to be taken
into consideration. However, it is adequate to take the deviation parameter
\textrm{k} as the only complex parameter without loss of generality ---$%
\mathrm{k}=\left\vert \mathrm{k}\right\vert e^{i\phi _{k}}$--- where $\phi
_{k}$\ is a $CP$ violating phase. In another vein, it is clear that the
neutrino mass matrix $m_{\nu }$\ in eq. (\ref{mn}) is not invariant under
the $\mu -\tau $\ symmetry\ transformation\ but still has the magic
symmetry, therefore, the neutrino matrix is diagonalized by the trimaximal
mixing matrix\ $\mathcal{U}_{TM_{2}}$ which can be\ parameterized as \cite%
{A9,A10,A11,A12,A13,A14,A15}%
\begin{equation}
\mathcal{U}_{TM_{2}}=\left(
\begin{array}{ccc}
\sqrt{\frac{2}{3}}\cos \theta & \frac{1}{\sqrt{3}} & \sqrt{\frac{2}{3}}\sin
\theta e^{-i\gamma } \\
-\frac{\cos \theta }{\sqrt{6}}-\frac{\sin \theta }{\sqrt{2}}e^{i\gamma } &
\frac{1}{\sqrt{3}} & \frac{\cos \theta }{\sqrt{2}}-\frac{\sin \theta }{\sqrt{%
6}}e^{-i\gamma } \\
-\frac{\cos \theta }{\sqrt{6}}+\frac{\sin \theta }{\sqrt{2}}e^{i\gamma } &
\frac{1}{\sqrt{3}} & -\frac{\cos \theta }{\sqrt{2}}-\frac{\sin \theta }{%
\sqrt{6}}e^{-i\gamma }%
\end{array}%
\right)  \label{TM}
\end{equation}%
The total mixing matrix, denoted as $\mathcal{U}_{\nu }$, can be expressed
as $\mathcal{U}_{\nu }=\mathcal{U}_{TM_{2}}.\mathcal{U}_{P}$ where\textrm{\ }%
$\mathcal{U}_{P}=diag(1,e^{i\frac{\alpha _{21}}{2}},e^{i\frac{\alpha _{31}}{2%
}})$\textrm{\ }is a diagonal matrix involving\ the Majorana $CP$ phases $%
\alpha _{21}$\ and $\alpha _{31}$. The two free parameters, $\theta $\ and $%
\gamma $, correspond to an arbitrary angle and phase, respectively, and they
are related to the observed neutrino mixing angles $\theta _{ij}$\ and the
Dirac $CP$\ phase $\delta _{CP}$. Indeed, the diagonalization of the
neutrino matrix (\ref{mn}) by the TM$_{\mathrm{2}}$\ mixing pattern induces
relations among these parameters as%
\begin{equation}
\tan 2\theta =\frac{\sqrt{3}\left\vert k\right\vert \sqrt{b^{2}\cos ^{2}\phi
_{k}+a^{2}\sin ^{2}\phi _{k}}}{2ab-b\left\vert k\right\vert \cos \phi _{k}}%
\quad ,\quad \tan \gamma =\frac{-a}{b}\tan \phi _{k}  \label{tmr}
\end{equation}%
Accordingly, the eigenmasses of the neutrino mass matrix $m_{\nu }$\ are as
follows%
\begin{eqnarray}
\left\vert m_{1}\right\vert &=&\frac{m_{0}}{\sqrt{(a-b)^{2}-\left\vert
\mathrm{k}\right\vert (a-b)\cos \phi _{k}+(\left\vert \mathrm{k}\right\vert
^{2}/4)}}\quad  \notag \\
\quad \left\vert m_{2}\right\vert &=&\frac{m_{0}}{\sqrt{(a+2b)^{2}+2\left%
\vert \mathrm{k}\right\vert (a+2b)\cos \phi _{k}+\left\vert \mathrm{k}%
\right\vert ^{2}}}  \label{mass} \\
\text{\ \ \ }\left\vert m_{3}\right\vert &=&\frac{m_{0}}{\sqrt{%
(a+b)^{2}-\left\vert \mathrm{k}\right\vert (a+b)\cos \phi _{k}+(\left\vert
\mathrm{k}\right\vert ^{2}/4)}}\text{\ }  \notag
\end{eqnarray}%
$\allowbreak $while the right handed neutrino masses $M_{1,2,3}$ are
expressed as%
\begin{eqnarray}
\left\vert M_{1}\right\vert &=&\Lambda \sqrt{(a-b)^{2}-\left\vert \mathrm{k}%
\right\vert (a-b)\cos \phi _{k}+(\left\vert \mathrm{k}\right\vert ^{2}/4)}
\notag \\
\left\vert M_{2}\right\vert &=&\Lambda \sqrt{(a+2b)^{2}+2\left\vert \mathrm{k%
}\right\vert (a+2b)\cos \phi _{k}+\left\vert \mathrm{k}\right\vert ^{2}}
\label{Maj} \\
\text{\ }\left\vert M_{3}\right\vert &=&\Lambda \sqrt{(a+b)^{2}-\left\vert
\mathrm{k}\right\vert (a+b)\cos \phi _{k}+(\left\vert \mathrm{k}\right\vert
^{2}/4)}  \notag
\end{eqnarray}

Considering that the obtained charged lepton mass matrix in Eq. (\ref{yl})
is diagonal, the lepton mixing results from the neutrino sector as $U_{PMNS}=%
\mathcal{U}_{\nu }=\mathcal{U}_{TM_{2}}$. Accordingly, By using the PDG
standard parametrization of the Pontecorvo-Maki-Nakagawa-Sakata (PMNS)
matrix \cite{B7}, we can derive expressions of the three neutrino mixing
angles $\theta _{ij}$ in terms of the trimaximal mixing parameters $\theta $%
\ and $\gamma $, we obtain%
\begin{equation}
\sin ^{2}\theta _{13}=\frac{2}{3}\sin ^{2}\theta ~,~\sin ^{2}\theta _{12}=%
\frac{1}{3-2\sin ^{2}\theta }~,~\sin ^{2}\theta _{23}=\frac{1}{2}-\frac{%
3\sin 2\theta }{2\sqrt{3}(3-2\sin ^{2}\theta )}\cos \gamma .  \label{angles}
\end{equation}

\subsection{Numerical analysis}

The neutrino sector in our model has five independent parameters namely $%
m_{0}$, $a$, $b$, $\mathrm{k}$ and $\phi _{k}$, whose allowed intervals can
be constrained by the numerical values of the neutrino oscillation parameters%
$.$ Moreover, due to the ambiguity in the sign of the atmospheric
mass-squared difference $\Delta m_{ij}^{2}$, our numerical investigation
encompasses two possible mass orderings; the normal hierarchy (\textbf{NH})
where $m_{1}<m_{2}<m_{3}$, and the inverted hierarchy (\textbf{IH}) where $%
m_{3}<m_{1}<m_{2}$. As a result, the diagonal mass matrix of the light
neutrinos can be rewritten in terms of the lightest neutrino mass $%
m_{1}(m_{3})$\ and the mass-squared differences as follows%
\begin{eqnarray*}
m_{\nu } &=&diag(m_{1},\sqrt{m_{1}^{2}+\Delta m_{21}^{2}},\sqrt{%
m_{1}^{2}+\Delta m_{31}^{2}})\text{ \ \ \ \ \ \ \ \ \ \ \ \ \ \ \ \ \
\textbf{For NH}} \\
m_{\nu } &=&diag(\sqrt{m_{3}^{2}-\Delta m_{32}^{2}-\Delta m_{21}^{2}},\sqrt{%
m_{3}^{2}-\Delta m_{32}^{2}},m_{3})\text{ \ \ \ \ \ \ \textbf{For IH}}
\end{eqnarray*}

\begin{figure}[h]
\begin{center}
\includegraphics[width=.43\textwidth]{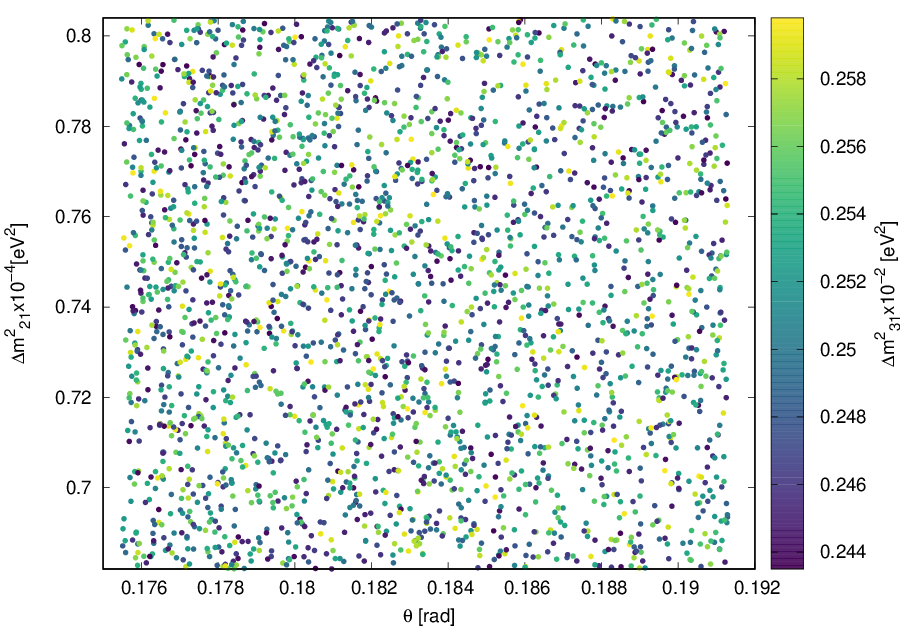}\includegraphics[width=.43%
\textwidth]{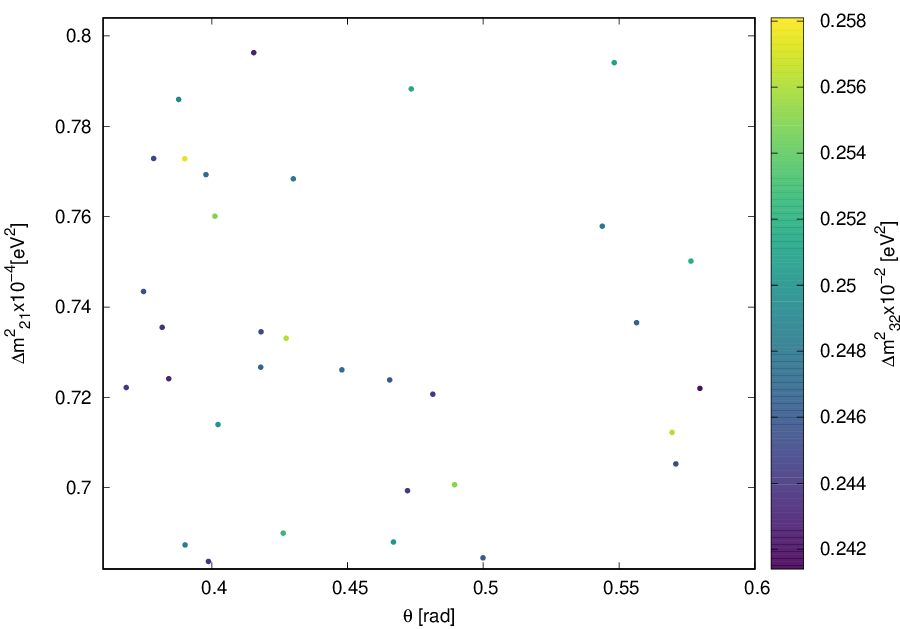}
\end{center}
\par
\vspace{-2em}
\caption{Correlations between the mass-squared differences $\Delta
m_{ij}^{2} $ and the parameter $\protect\theta $. The color palette
corresponds to $\Delta m_{31}^{2}$ for NH (left panel) and to $\Delta
m_{32}^{2}$ for IH (right panel).}
\label{1}
\end{figure}

\begin{figure}[h]
\begin{center}
\hspace{2.5em} \includegraphics[width=.45\textwidth]{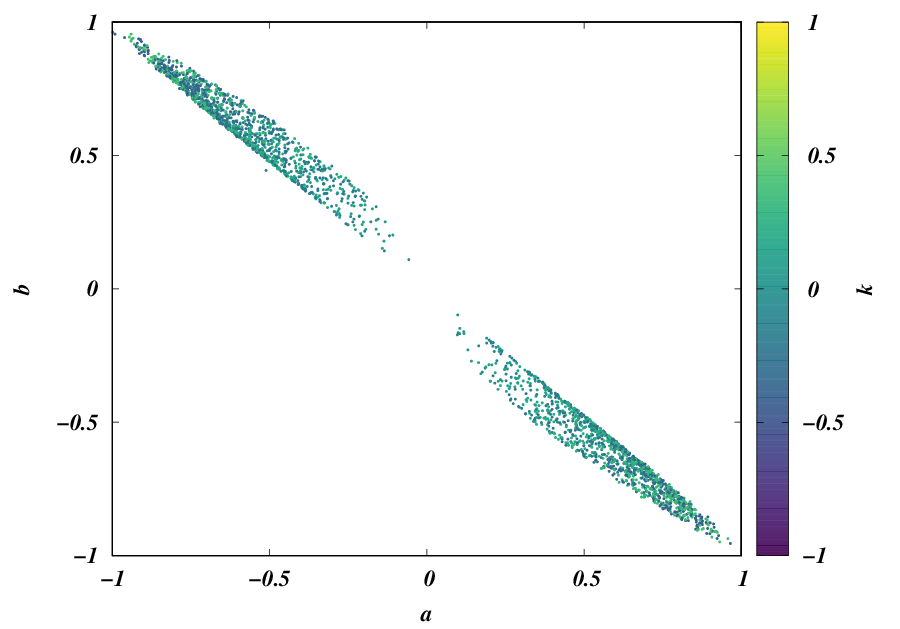} %
\includegraphics[width=.45\textwidth]{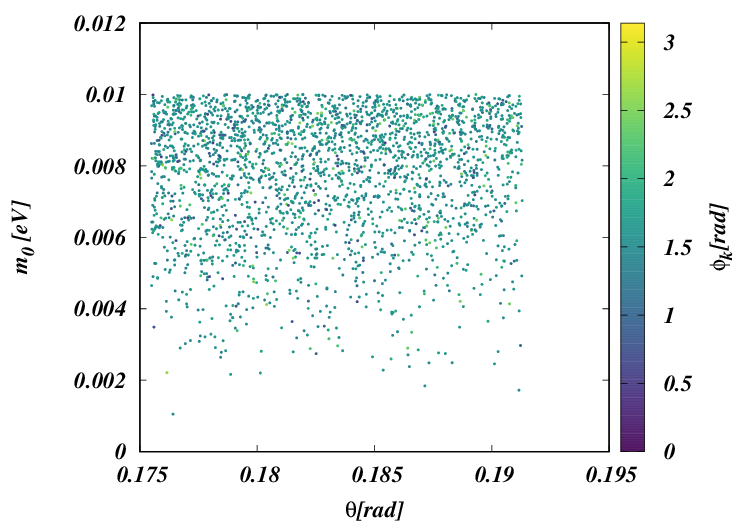}
\end{center}
\caption{The model parameters constrained by $3\protect\sigma $ ranges of
the oscillation parameters. the left panel shows the correlation among the
parameters $a$, $b$ and \textrm{k}. The right panel shows the correlation
among $m_{0}$, $\protect\theta $\ and $\protect\phi _{k}$.}
\label{2}
\end{figure}

\begin{figure}[h]
\centering\includegraphics[width=.42\textwidth]{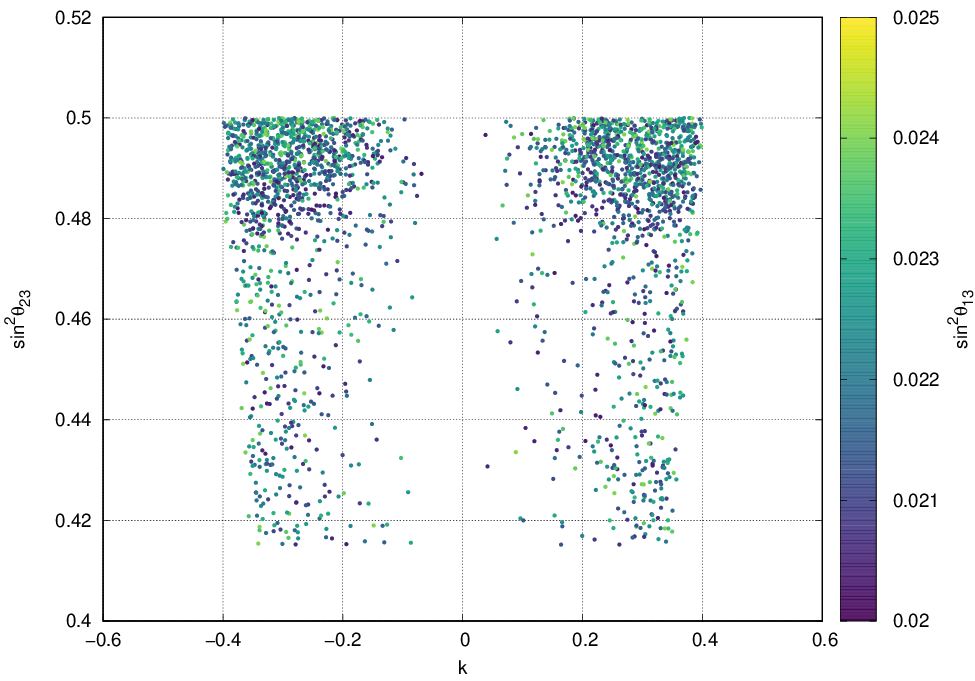}
\caption{Scatter plot on the plane of \textrm{k} and $\sin ^{2}\protect%
\theta _{23}$ with the palette corresponds to $\sin ^{2}\protect\theta _{13}$%
.}
\label{f3}
\end{figure}

Considering the $3\sigma $ experimental ranges of $\Delta m_{ij}^{2}$
reported in Table (\ref{t0}) along with the eigenmasses in Eq. (\ref{mass})
and the relation between $\theta $ and the free parameters in Eq. (\ref{tmr}%
), we plot in Figure (\ref{1}) the correlation between $\Delta m_{ij}^{2}$\
and the trimaximal mixing parameter $\theta $ for both hierarchies. In order
to satisfy the $3\sigma $ experimental intervals of the mass-squared
differences $\Delta m_{ij}^{2}$\ for both hierarchies, we find that the
parameter $\theta $ lies in the interval%
\begin{equation}
0.175\lesssim \theta \text{[rad]}\lesssim 0.191\text{ (NH) }~,\text{ }%
~0.368\lesssim \theta \text{[rad]}\lesssim 0.579\text{ (IH)}
\end{equation}%
Based on first and the second relations in Eq. (\ref{angles}), we conclude
that the obtained interval of $\theta $ in the case of NH is consistent with
the $3\sigma $ experimental ranges of the reactor angle $\theta _{13}$ and
atmospheric angle $\theta _{23}$. However, in the case of IH, the values of $%
\theta $ acquired are too large to be consistent with the experimental
values of the mixing angles $\theta _{13}$ and $\theta _{23}$. As a result,
the inverted hierarchy for the neutrino mass spectrum is excluded within our
model. To constrain the model parameter space in the NH scheme, we employ
the $3\sigma $ experimental ranges of the neutrino oscillation observables $%
\Delta m_{ij}^{2}$, $\sin \theta _{ij}$ and $\delta _{CP}$ as input
parameters. By using Eqs. (\ref{mass}) and (\ref{angles}) we present a
correlation plot among the parameters $a$, $b$\ and \textrm{k} in the left
panel of Figure (\ref{2}). We find that their allowed ranges are%
\begin{equation}
-0.9987\lesssim a\lesssim 0.9655\quad ,\quad -0.9541\lesssim b\lesssim
0.9613\quad ,\quad -0.4018\lesssim \mathrm{k}\lesssim 0.3985  \label{par}
\end{equation}%
Given the $3\sigma $ ranges of the neutrino oscillation data, we plot in the
right panel of Figure (\ref{2}) the parameter $m_{0}$ as a function of $%
\theta $ whereas the color map shows the values of the phase $\phi _{k}$. We
find that the constrained ranges of the parameters $m_{0}$ and $\phi _{k}$
are given as follows%
\begin{equation}
0.544\lesssim \phi _{k}\text{[rad]}\lesssim 2.629\qquad \text{and}\qquad
0.001\lesssim m_{0}\text{[eV]}\lesssim 0.01
\end{equation}%
Moreover, we plot in Figure (\ref{f3}) the correlation among $\sin
^{2}\theta _{23}$, $\sin ^{2}\theta _{13}$\ and the deviation parameter $%
\mathrm{k}$. We observe that the value $\mathrm{k}=0$ and the values around
it are not allowed, and this yields an important prediction of the current
model where it ensures a nonzero value of the reactor angle $\theta
_{13}\neq 0$\ and a non maximal value of the atmospheric angle $\theta
_{23}\neq \frac{\pi }{4}$. We find that the allowed region of $\sin
^{2}\theta _{13}$ lies in the range $[0.0203,0.0240]$ while the allowed
region of $\sin ^{2}\theta _{23}$ favors the second octant region $%
[0.415,0.499]$ which could be measured with more precision by the future
experiments.
Taking into account the non vanishing value of the reactor angle $\theta
_{13}$, possible $CP$ violating effects in neutrino oscillations can be
generated from nonzero value of the Dirac phase $\delta _{CP}$. The
magnitude of these effects is estimated by the Jarlskog invariant parameter
defined as\ $J_{CP}=Im(\mathcal{U}_{e1}\mathcal{U}_{\mu 1}^{\ast }\mathcal{U}%
_{\mu 2}\mathcal{U}_{e2}^{\ast })$\ \cite{B7}. In the standard
parametrization of the PMNS mixing matrix, this parameter is expressed in
terms of the three mixing angles and the Dirac $CP$ phase as follows \cite%
{B7}%
\begin{equation}
J_{CP}=\frac{1}{8}\sin 2\theta _{12}\sin 2\theta _{13}\sin 2\theta _{23}\cos
\theta _{13}\sin \delta _{CP}  \label{JCP}
\end{equation}%
\begin{figure}[h]
\begin{center}
\includegraphics[width=.45\textwidth]{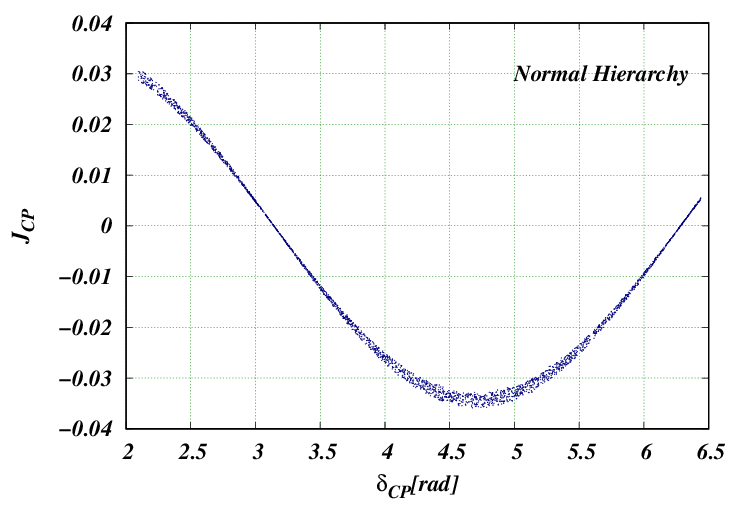} \includegraphics[width=.45%
\textwidth]{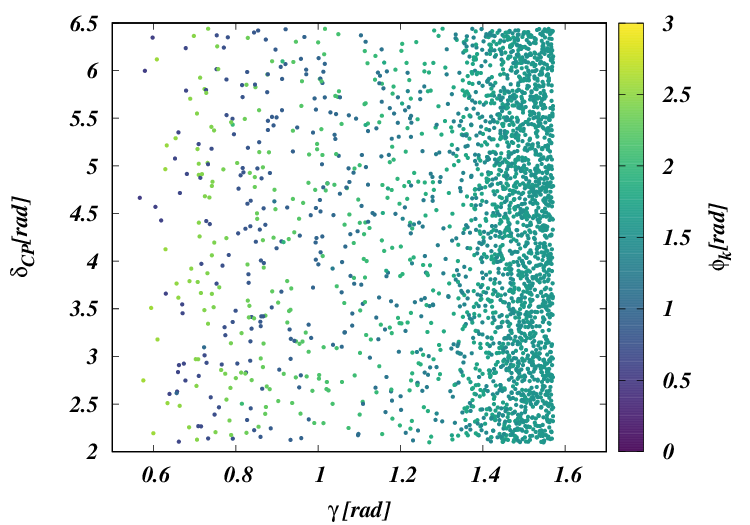}
\end{center}
\caption{Left panel: $J_{CP}$ as a function of the Dirac $CP$ phase $\protect%
\delta _{CP}$. Right panel: Correlation among the parameters $\protect\delta %
_{CP}$, $\protect\gamma $ and $\protect\phi _{k}$.}
\label{f4}
\end{figure}
Using the $3\sigma $ range of $\delta _{CP}$\ and the restricted ranges of
the mixing angles $\sin ^{2}\theta _{ij}$, we plot in the left panel of
Figure (\ref{f4}), the correlation between $J_{CP}$ and $\delta _{CP}$.
Furthermore, by combining the standard parametrization of the lepton mixing
matrix and Eq. (\ref{TM}), the Jarlskog invariant is reduced to $J_{CP}=%
\frac{\sin 2\theta \sin \gamma }{6\sqrt{3}}$. Accordingly, by identifying
the two expressions of the $J_{CP}$ parameter, we plot in the right panel of
Figure (\ref{f4}) the correlation among $\delta _{CP}$, $\gamma $ and $\phi
_{k}$. We find that the Dirac $CP$ violating phase is not restricted
compared to its $3\sigma $\ range including the conserving values $\delta
_{CP}=\pi ,2\pi $. We find also that the Jarlskog invariant parameter falls
in the range $-0.0359\lesssim J_{CP}\lesssim 0.0305$ whereas the constrained
interval of the trimaximal mixing parameter $\gamma $ is given by
\begin{equation}
0.566\lesssim \gamma \text{[rad]}\lesssim 1.570  \label{gama}
\end{equation}

\section{Predictions on the absolute neutrino mass scale}

\label{sec3} The neutrino oscillation experiments are sensitive to the
neutrino mixing angles $\theta _{ij}$ and to the mass-squared differences $%
\Delta m_{ij}$ as reported in Table (\ref{t0}), however, they are not
capable to provide information on the absolute neutrino mass scale. Direct
measurements of the absolute mass scale is one of the most important
purposes of the next-generation neutrino experiments. In this respect, the
absolute mass scale can be determined from non oscillation methods, using
Tritium beta decay \cite{B8}, neutrinoless double beta decay \cite{B9}, and
cosmological observations \cite{B10,B11}.

\subsection{Neutrino masses from cosmology}

\begin{figure}[h]
\begin{center}
\includegraphics[width=.42\textwidth]{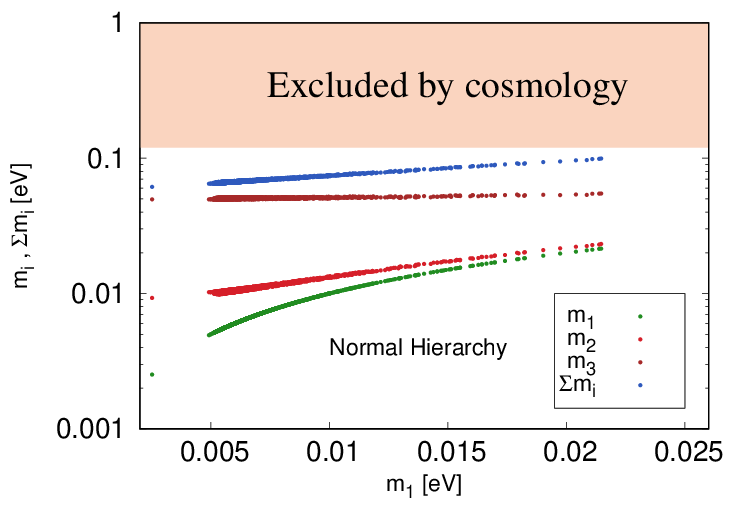}\includegraphics[width=.42%
\textwidth]{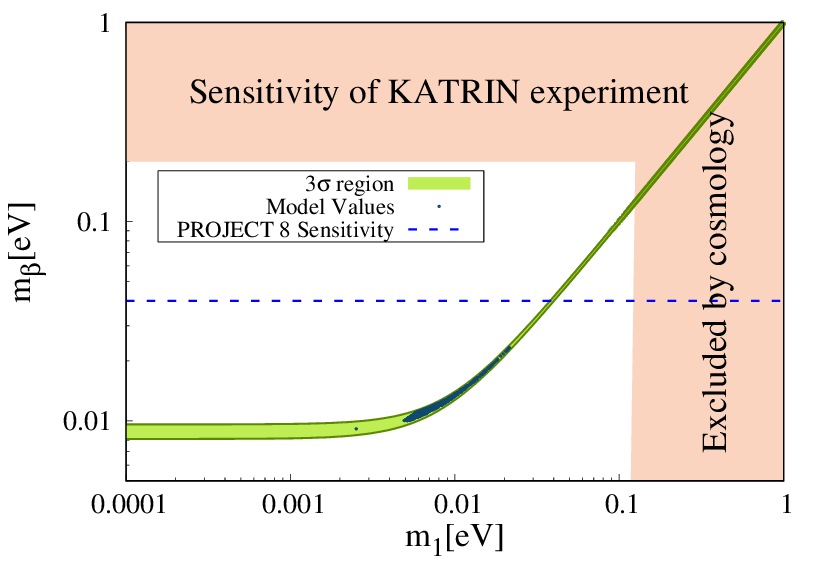} \includegraphics[width=.42\textwidth]{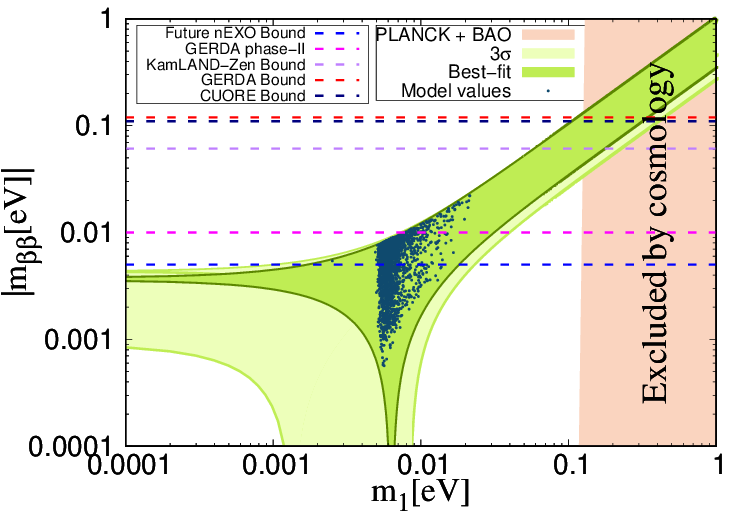}
\end{center}
\par
\vspace{-2em}
\caption{Top left panel: Prediction for the absolute neutrino masses $m_{i}$
and their sum $\sum m_{i}$ as a function of $m_{1}$. The horizontal region
is disfavored by Planck+BAO. Top right panel: $m_{\protect\beta }$ as a
function of the lightest neutrino mass $m_{1}$. The vertical region is
disfavored by Planck+BAO while the horizontal bound is the limit on $m_{%
\protect\beta }$ from KATRIN collaboration. Bottom panel: The effective
Majorana mass $\left\vert m_{\protect\beta \protect\beta }\right\vert $\ as
a function of the lightest neutrino mass $m_{1}$. The vertical bound
represents the upper limit on the sum of the three light neutrino masses.}
\label{f5}
\end{figure}

Cosmological observations could further constrain neutrino masses by
providing information on the sum of all neutrino masses. The Planck
collaboration analysis ---which is based on the $\Lambda $CDM cosmological
model--- including data on baryon acoustic oscillations (BAO) provided an
upper limit on the total neutrino mass of $\sum m_{i}<0.12$ $\mathrm{eV}$\
at 95\% C.L \cite{B10}\textrm{. }Taking into account this sum and
incorporating the $3\sigma $\ ranges of the neutrino oscillation parameters $%
\theta _{ij}$\ and $\Delta m_{ij}^{2}$ as well as the obtained intervals of
the model parameters, we plot in the top left panel of Figure (\ref{f5}) the
lightest neutrino mass $m_{1}$ as a function of the three neutrino masses $%
m_{i=1,2,3}$, and their sum $\sum m_{i}$. We observe that the neutrino
masses $m_{1}$ and $m_{2}$ lie in the intervals $0.002516\lesssim
m_{1}\left( \mathrm{eV}\right) \lesssim 0.021351$ and\ $0.009859\lesssim
m_{2}\left( \mathrm{eV}\right) \lesssim 0.023014$ while the values of $m_{3}$
lies in a narrow region $0.049619\lesssim m_{3}\left( \mathrm{eV}\right)
\lesssim 0.054751$. Moreover, we find that the obtained range of the sum of
all three absolute neutrino masses is given by\textrm{\ }$0.064744\lesssim
\sum m_{i}\left( \mathrm{eV}\right) \lesssim 0.098468$, which satisfies the
cosmological bound on the sum of light neutrino masses. Notably, the
predicted values around the lower bound of $\sum m_{i}\sim 0.062$ $\mathrm{eV%
}$ hold particular interest in upcoming experiments, like CORE+BAO. These
experiments are expected to provide additional cosmological data with the
potential to reach a sensitivity of around $0.062$ $\mathrm{eV}$ on the sum
of the three light neutrino masses\textrm{\ }\cite{B12}\textrm{.}

\subsection{Search for the neutrinoless double beta decay}

The nature of neutrinos ---Dirac or Majorana---\ is one of the most
outstanding questions in neutrino physics, and the neutrinoless double beta (%
$0\nu \beta \beta $) decay is the only known process capable of testing the
intrinsic nature of neutrinos \cite{B9}\textrm{.} If the $0\nu \beta \beta $
decay is observed, it would imply a violation of lepton number \emph{L} by
two units and provide strong evidence that neutrinos are Majorana fermions.
Furthermore, this process can also probe the absolute neutrino mass scale by
measuring the effective Majorana mass of the electron neutrino. This later
is defined as $\left\vert m_{\beta \beta }\right\vert =\left\vert
\sum_{i}U_{ei}^{2}m_{i}\right\vert $ where $U_{ei}$ are the entries of
lepton mixing matrix and correspond to its first row while $m_{i}$ are the
three neutrino mass eigenvalues. For our theoretical framework, the
effective Majorana mass $\left\vert m_{\beta \beta }\right\vert $\ can be
rewritten in terms\textrm{\ }of the $\mathcal{U}_{TM_{2}}$\ mixing matrix
parameters and the lightest neutrino mass $m_{1}$\ as%
\begin{equation}
\left\vert m_{\beta \beta }\right\vert =\left\vert \frac{2m_{1}}{3}\cos
^{2}\theta +\frac{1}{3}\sqrt{m_{1}^{2}+\Delta m_{21}^{2}}e^{\frac{i}{2}%
\alpha _{21}}+\frac{2}{3}\sin ^{2}\theta \sqrt{m_{1}^{2}+\Delta m_{31}^{2}}%
e^{\frac{i}{2}(\alpha _{31}-2\gamma )}\right\vert
\end{equation}%
Taking into account the $3\sigma $ ranges of the oscillation parameters $%
\theta _{ij}$, $\Delta m_{ij}^{2}$ as well as the restricted intervals of $%
\theta $, $\gamma $ and $m_{1}$, we show in the top right panel of Figure (%
\ref{f5}) the correlation between $\left\vert m_{\beta \beta }\right\vert $\
and the lightest neutrino mass $m_{1}$. The Majorana phases $\alpha _{21}$
and $\alpha _{31}$ are randomly varied within the interval $[0\rightarrow
2\pi ]$. The horizontal dashed lines represent the limits on $\left\vert
m_{\beta \beta }\right\vert $ from current experiments on $0\nu \beta \beta $
decay and the vertical bound is disfavored by the Planck+BAO data. Our
findings reveal that the range of $\left\vert m_{\beta \beta }\right\vert $
falls within $0.000567\lesssim \left\vert m_{\beta \beta }\right\vert \left(
\mathrm{eV}\right) \lesssim 0.022121$. Notably, this region is below the
experimental limits set by KamLAND-Zen \cite{B13}, CUORE \cite{B14} and
GERDA \cite{B15} experiments, which impose constraints on $\left\vert
m_{\beta \beta }\right\vert <(0.061-0.165)$ \textrm{eV}, $\left\vert
m_{\beta \beta }\right\vert <(0.075-0.35)$ \textrm{eV} and $\left\vert
m_{\beta \beta }\right\vert <(0.104-0.228)$ \textrm{eV}, respectively.
Furthermore, upcoming experiments such as GERDA Phase II \cite{B16},\textrm{%
\ }nEXO \cite{B17}, CUPID \cite{B18} and SNO+-II \cite{B19} have the
potential to experimentally probe this range of values. These experiments
aim to achieve sensitivities in the range of $m_{\beta \beta }\sim \left(
0.01-0.02\right) $ \textrm{eV}, $\left\vert m_{\beta \beta }\right\vert \sim
\left( 0.006-0.017\right) $ \textrm{eV}, $\left\vert m_{\beta \beta
}\right\vert \sim \left( 0.008-0.022\right) $ \textrm{eV} and $\left\vert
m_{\beta \beta }\right\vert \sim \left( 0.02-0.07\right) $ \textrm{eV},
respectively. Therefore, these upcoming experiments hold the potential to
verify and confirm the predicted values of $\left\vert m_{\beta \beta
}\right\vert $ within their respective sensitivities, thereby serving as
crucial tests for the model.

\subsection{Direct determination of the neutrino mass by kinematics}

Tritium beta decay experiments, which measure the end-point electron
spectrum, provide the most sensitive method of determining the mass of the
electron neutrino. These experiments constrain the effective electron
neutrino mass, denoted as $m_{\beta }$, which can be expressed as $m_{\beta
}=\left( \sum_{i}\left\vert U_{ei}\right\vert ^{2}m_{i}^{2}\right) ^{1/2}$.
By relating this mass in terms of the\textrm{\ }TM$_{\text{\textrm{2}}}$
parameters and the lightest neutrino mass $m_{1}$, we obtain the following
relation%
\begin{equation}
m_{\beta }=\left( \frac{2m_{1}^{2}}{3}\cos ^{2}\theta +\frac{1}{3}%
(m_{1}^{2}+\Delta m_{21}^{2})+\frac{2}{3}\sin ^{2}\theta (m_{1}^{2}+\Delta
m_{31}^{2})e^{-i\gamma )}\right) ^{1/2}
\end{equation}%
With respect to the constrained ranges of the relevant parameters in the
expression of $m_{\beta }$, we show in the bottom panel of Figure (\ref{f5})
the correlation between $m_{\beta }$ and $m_{1}$\ where we find that $%
m_{\beta }$ falls in the region $0.0091213\lesssim m_{\beta }\left( \mathrm{%
eV}\right) \lesssim 0.023347$. These predicted values are bellow the present
upper limit $m_{\beta }<1.1$ \textrm{eV} provided by KATRIN experiment at
90\% CL \cite{B20}. Furthermore, these values are significantly smaller
compared to the forecasted sensitivities coming from the future $\beta $%
-decay experiments such as KATRIN ($\sim 0.2$ \textrm{eV}) \cite{B21},
Project 8 ($\sim 0.04$ \textrm{eV}) \cite{B22} and HOLMES ($\sim 0.1$
\textrm{eV}) \cite{B23}. If the effective electron neutrino mass is measured
by any of these experiments, it would lead to the exclusion of the current
model. Conversely, if none of these experiments are able to measure the
effective electron neutrino mass $m_{\beta }$, the model's predictions for $%
m_{\beta }$ could be investigated by future experiments aiming to achieve
improved sensitivities around $0.01$ \textrm{eV}.

\section{ Leptogenesis}

\label{sec4}

In this section, we shed light on the origin of the baryon asymmetry of the
universe in our setup where we perform numerical analysis to investigate the
model's implications for leptogenesis.

\subsection{Baryogenesis through unflavored leptogenesis}

The leptogenesis mechanism, first proposed by Fukugita and Yanagida \cite%
{A43}, is one of the most attractive scenarios that can explain the origin
of the baryon asymmetry of the universe. In this respect, the lepton
asymmetry is generated by the decay of heavy right-handed neutrinos ---which
are naturally present in the type I seesaw framework--- into lepton and
Higgs particles. This created asymmetry is then transferred into the baryon
sector through the so-called sphaleron processes. In this section, we
compute the baryon asymmetry parameter within our framework through the
decay of heavy singlet neutrinos $N_{i}$ pursuing the following approaches

\begin{itemize}
\item The lepton asymmetry is generated through the out-of-equilibrium decay
of all Majorana neutrinos, taking into consideration that the right-handed
neutrino mass spectrum is not strongly hierarchical.

\item The baryon asymmetry parameter is calculated within \emph{the} \emph{%
unflavored approximation}, where the generation of lepton asymmetry occurs
at a temperature of the universe $T\sim M_{i}\gtrsim 5\times 10^{11}(1+\tan
^{2}\beta )$ \textrm{GeV}. This approximation neglects flavor effects,
treating all charged lepton flavors as indistinguishable.
\end{itemize}

\begin{figure}[h]
\begin{center}
\includegraphics[width=.45\textwidth]{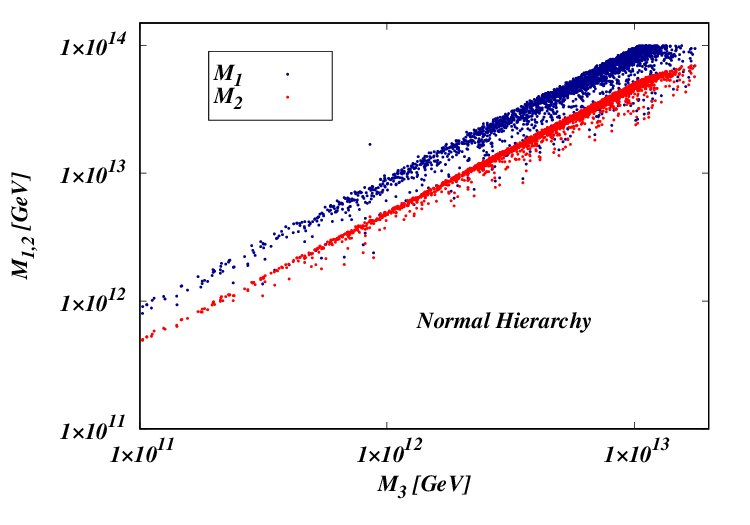}
\end{center}
\par
\vspace{-2em}
\caption{Majorana masses $M_{1}$ and $M_{2}$ as a function of the lightest
Majorana mass $M_{3}$.}
\label{11}
\end{figure}
To proceed with the calculation of $Y_{B}$ in our scenario, it is necessary
to establish the RH neutrino mass spectrum and explain how the leptogenesis
can be investigated within the unflavored regime. To achieve this, by using
the expressions of the RH neutrino masses in eq. (\ref{Maj}), we show in
Figure (\ref{11}) the correlation between the RH neutrino masses $M_{1,2}$
and the lightest Majorana mass $M_{3}$. We observe that the RH neutrino mass
spectrum in our model does not exhibit strong hierarchy ($M_{1}\sim
M_{2}\sim 3M_{3}$). Moreover, given that the majority of data points
consistent with the observed neutrino oscillations lie above the limit $%
M_{i}\gtrsim 10^{12}$, the use of the unflavored approximation is favorable
in our analysis. This approximation holds as long as the masses of the RH
neutrinos satisfy the limit $T\sim M_{i}\gtrsim 5\times 10^{11}(1+\tan
^{2}\beta )$ \textrm{GeV}. Specifically, for small values of $\tan \beta =3$%
, the lower limit is estimated to be $T\sim M_{i}\gtrsim $ $5.0\times
10^{12} $ GeV.\newline
To estimate the BAU produced in our model, we recall that the cosmological
baryon asymmetry $Y_{B}$ can be expressed as the ratio between the net
baryon number density and the entropy density $s$ of the universe as%
\begin{equation}
Y_{B}=\frac{n_{B}-n_{\overline{B}}}{s}
\end{equation}%
where $n_{B}$ and $n_{\overline{B}}$\ are the number densities of baryons
and anti-baryons respectively. The observed baryon asymmetry of the universe
from Planck satellite is given by $Y_{B}=(8.72\pm 0.08)\times 10^{-11}$\ at
the $1\sigma $ level \cite{D1}. Taking into consideration the significant
contribution of each RH neutrino to the baryon asymmetry $Y_{B}$, it can be
expressed in the following general form%
\begin{equation}
Y_{B}=\sum_{i=1}^{3}Y_{B_{i}}  \label{YB}
\end{equation}%
where the quantities $Y_{B_{i}}$ correspond to the part of the baryon
asymmetry produced by the \textit{i}th RH neutrino. This can be formally
expressed as\emph{\ }\cite{D2}%
\begin{equation}
Y_{B_{i}}=-2c_{s}\frac{n_{N_{i}}}{s}\epsilon _{i}\eta _{ii}
\end{equation}%
where $\epsilon _{i}$ is the $CP$ asymmetry parameter produced in the decay
of $N_{i}^{c}$ and $\eta _{ii}$\ are the efficiency factors describing the
fraction of the $CP$ asymmetry that survives the washout by inverse decays
and scattering processes, $c_{s}$ is the fraction of the $B-L$ asymmetry
converted into baryon asymmetry by sphalerons ($c_{s}=32/92$ in the MSSM)
and $\frac{n_{N_{i}}}{s}$ is the number density of right-handed neutrinos
normalized to the entropy density; it is defined as \cite{D3}%
\begin{equation}
\frac{n_{N_{i}}}{s}=\frac{135\zeta (3)}{4\pi ^{4}g_{\ast }}
\end{equation}%
\newline
where $\zeta (3)$\ is the Riemann zeta function and $g_{\ast }$ is the
number of spin-degrees of freedom in thermal equilibrium; in the MSSM we
have $g_{\ast }=228.75$ \cite{A44,D4,D5}. Accordingly, the baryon asymmetry
produced by the \textit{i}th RH neutrino can be approximated as%
\begin{equation}
Y_{B_{i}}\simeq -1.26\times 10^{-3}\epsilon _{i}\eta _{ii}  \label{YBi}
\end{equation}

\subsection{Estimating $CP$ asymmetry}

The baryon asymmetry $Y_{B_{i}}$ is mainly related to the two important
quantities $\epsilon _{i}$ and $\eta _{ii}$ which are model dependent. When
dealing with SUSY models, the $CP$ asymmetry parameter $\epsilon _{i}$ can
be explicitly expressed in the unflavored approximation as%
\begin{equation}
\epsilon _{N_{i}}=\frac{1}{8\pi }\sum_{j=1,2}\frac{Im\left[ \left( \mathcal{Y%
}_{\nu }\mathcal{Y}_{\nu }^{\dagger }\right) _{j3}^{2}\right] }{\left(
\mathcal{Y}_{\nu }\mathcal{Y}_{\nu }^{\dagger }\right) _{33}}f\left( \frac{%
M_{j}}{M_{i}}\right)  \label{epsilon}
\end{equation}%
where the loop function is defined as $f\left( x\right) =\sqrt{x}\left(
1-\left( 1+x\right) \ln \left[ \left( 1+x\right) /x\right] \right) $ and%
\emph{\ }$\mathcal{Y}_{\nu }$\emph{\ }is the neutrino Yukawa coupling matrix
in the basis where the Majorana mass matrix is diagonal. It is clear from
eq. (\ref{epsilon}) that the non vanishing $CP$\ asymmetry parameter $%
\epsilon _{N_{i}}$ requires the off-diagonal entries of the product $%
\mathcal{Y}_{\nu }\mathcal{Y}_{\nu }^{\dagger }$ to be simultaneously
nonzero and complex. However, considering the Dirac Yukawa matrix in eq. (%
\ref{DM}), we find that the product $\mathcal{Y}_{\nu }\mathcal{Y}_{\nu
}^{\dagger }$ is proportional to the identity matrix. Therefore, the lepton
asymmetry $\epsilon _{N_{i}}$ vanishes and the baryon asymmetry can not be
generated at LO in our model. Therefore, to ensure a non vanishing baryon
asymmetry $Y_{B}$, higher order corrections to the Dirac mass matrix must be
considered\footnote{%
In models with a twofold degenerate Dirac mass matrix $m_{D}=diag(a,b,b)$,
the lepton asymmetries $\varepsilon _{N_{i}}$ are not vanishing and the
correct amount of lepton asymmetry depends on $\varepsilon _{N_{i}}\sim
(\left\vert b\right\vert ^{2}-\left\vert a\right\vert ^{2})^{2}$. A detailed
calculation of the baryon asymmetry of the universe in the case of Dirac
neutrino mass matrix $M_{D}$ with two degenerate eigenvalues is performed in
Ref. \cite{Y1}. This study is performed\ within $Z_{2}$ model \cite{Y2} and $%
D_{4}$ model \cite{A36}.}.\newline
To generate a sufficiently large baryon asymmetry $Y_{B}$,\ we introduce a
new flavon field $\omega $ which transforms as $1_{+-}$ under $D_{4}$ with
zero $U(1)$ charge. The latter gives rise the higher order correction $%
\delta W_{D}$ given by\footnote{%
We should mention that the Dirac Yukawa couplings at NLO of the form $\frac{%
\lambda _{ij}}{\Lambda }N_{i}^{c}L_{j}H_{u}\digamma _{1}$\ and $\frac{%
\lambda _{ij}}{\Lambda ^{2}}N_{i}^{c}L_{j}H_{u}\digamma _{1}\digamma _{2}$\
where $\digamma _{1},\digamma _{2}=\rho _{1},\rho _{2},\rho _{3},\eta
,\sigma ,\phi ,\chi ,\psi $ are forbidden by $D_{4}\times U(1)$ flavor
symmetry. Moreover, the contribution of 7-dimensional operators of the form $%
\frac{\lambda _{ij}}{\Lambda ^{3}}N_{i}^{c}L_{j}H_{u}\digamma _{1}\digamma
_{2}\digamma _{3}$ are expected to be too small and consequently the $CP$
asymmetry parameter $\epsilon _{N_{i}}$ is strongly suppressed.}%
\begin{equation}
\delta W_{D}=\frac{\lambda _{9}}{\Lambda }N_{3,2}^{c}L_{\mu ,\tau
}H_{u}\omega
\end{equation}%
where $\lambda _{9}$ is a complex coupling constant $\lambda _{9}=\left\vert
\lambda _{9}\right\vert e^{i\phi _{\omega }}$. When the singlet flavon field
$\omega $ acquires its VEV as $\left\langle \omega \right\rangle =\upsilon
_{\omega }$, the total Dirac Yukawa mass matrix becomes
\begin{equation}
\mathcal{Y}_{D}=Y_{D}+\delta Y_{D}=\frac{m_{D}}{\upsilon _{u}}+\delta
Y_{D}=\left(
\begin{array}{ccc}
\lambda _{1} & 0 & 0 \\
0 & \lambda _{1} & 0 \\
0 & 0 & \lambda _{1}%
\end{array}%
\right) +he^{i\phi _{\omega }}\left(
\begin{array}{ccc}
0 & 0 & 0 \\
0 & 0 & 1 \\
0 & 1 & 0%
\end{array}%
\right)
\end{equation}%
where $h=\frac{\left\vert \lambda _{9}\right\vert \upsilon _{\omega }}{%
\Lambda }$ is a free parameter which must be small in order to produce the
correct BAU\footnote{%
Notice that the contribution $\delta W_{D}$ in the Dirac mass matrix is
small compared to the leading order contribution and will not have provide
any significant effect on the results obtained regarding the neutrino sector.%
}. The total Yukawa neutrino mass matrix $\mathcal{Y}_{\nu }$ relevant for
the calculation of $\epsilon _{N_{i}}$ is defined as $\mathcal{Y}_{\nu }=%
\mathcal{U}_{\nu }^{\dagger }\mathcal{Y}_{D}$. Thus, the analytic
expressions for the $CP$ asymmetry parameters $\epsilon _{N_{i}}$ generated
in the decays of RH neutrinos $N_{i}$ are given approximately by

\begin{eqnarray}
\epsilon _{N_{1}} &=&\frac{h^{2}}{9\pi }\cos ^{2}\phi _{\omega }\left[
\left( \cos ^{2}\left( \theta \right) \sin ^{2}\left( \frac{\alpha
_{21}+4\phi _{\omega }}{2}\right) \right) f\left( \frac{\tilde{m}_{2}}{%
\tilde{m}_{1}}\right) +\left( 2\sin ^{2}\left( \frac{\alpha _{21}-2\gamma }{2%
}\right) \sin ^{2}(2\theta )\right) f\left( \frac{\tilde{m}_{3}}{\tilde{m}%
_{1}}\right) \right]  \notag \\
\epsilon _{N_{2}} &=&\frac{h^{2}}{9\pi }\cos ^{2}\phi _{\omega }\left[
\left( \cos ^{2}(\theta )\sin ^{2}\left( \frac{\alpha _{21}}{2}\right)
\right) f\left( \frac{\tilde{m}_{1}}{\tilde{m}_{2}}\right) +\left( \sin
^{2}\left( \frac{\alpha _{21}-\alpha _{31}+2\gamma }{2}\right) \sin
^{2}\left( \theta \right) \right) f\left( \frac{\tilde{m}_{3}}{\tilde{m}_{2}}%
\right) \right]  \label{ep123} \\
\epsilon _{N_{3}} &=&\frac{h^{2}}{9\pi }\cos ^{2}\phi _{\omega }\left[
\left( 2\sin ^{2}(2\theta )\sin ^{2}\left( \frac{\alpha _{31}-2\sigma }{2}%
\right) \right) f\left( \frac{\tilde{m}_{1}}{\tilde{m}_{3}}\right) +\left(
\sin ^{2}\left( \frac{\alpha _{21}-\alpha _{31}+2\gamma }{2}\right) \sin
^{2}\left( \theta \right) \right) f\left( \frac{\tilde{m}_{2}}{\tilde{m}_{3}}%
\right) \right]  \notag
\end{eqnarray}%
where $\tilde{m}_{i}$ are the washout mass parameters expressed as $\tilde{m}%
_{i}=\upsilon _{u}^{2}\frac{\left( \mathcal{Y}_{\nu }\mathcal{Y}_{\nu
}^{\dagger }\right) _{ii}}{M_{i}}$. Now we turn to discuss the the
efficiency factors $\eta _{ii}$ necessary for estimating the baryon
asymmetry $Y_{B}$. In general, its computation requires numerical solution
of the Boltzmann equations. However, as the RH neutrino masses are taken in
our scenario to be smaller than $10^{14}$ $\mathrm{GeV}$, possible washout
effects from $\Delta L=2$ scattering processes are out of equilibrium. As a
result, the efficiency factors $\eta _{ii}$ can be approximated as a
function of the washout mass parameter $\tilde{m}_{i}$ as \cite{D6}%
\begin{equation}
\eta _{ii}\approx \left( \frac{3.3\times 10^{-3}\text{\textrm{eV}}}{\tilde{m}%
_{i}}+\left( \frac{\tilde{m}_{i}}{0.55\times 10^{-3}\text{\textrm{eV}}}%
\right) ^{1.16}\right) ^{-1}
\end{equation}%
Note here that the smallness of the parameter $h<<\lambda _{1}$ implies that
the washout mass parameters and the neutrino masses become approximately
equal $\tilde{m}_{i}\approx m_{i}$. As a result, the efficiency factors are
functions of the light neutrino masses\ $m_{i}$\ as $\eta _{ii}(\tilde{m}%
_{i})\approx \eta _{ii}(m_{i})$.\newline
From the previous formulation of the $CP$\ asymmetry parameters\ $\epsilon
_{N_{i}}$\ in eq. (\ref{ep123}) and the efficiency factors $\eta _{ii}$\ as
shown above, it can be seen that the the total baryon asymmetry $Y_{B}$\
depends mainly on the parameters resulting from the correction $\delta W_{D}$%
\ in the Dirac mass matrix ---namely the parameter $h$\ and the phase $\phi
_{\omega }$,\ which serves as a new source of $CP$ violation--- as well as
on the trimaximal parameters $\theta $\ and $\gamma $, the light neutrino
masses $m_{i}$\ and the Majorana phases $\alpha _{31}$\ and $\alpha _{21}$.

\subsection{Numerical analysis}

\begin{figure}[h]
\begin{center}
\includegraphics[width=.43\textwidth]{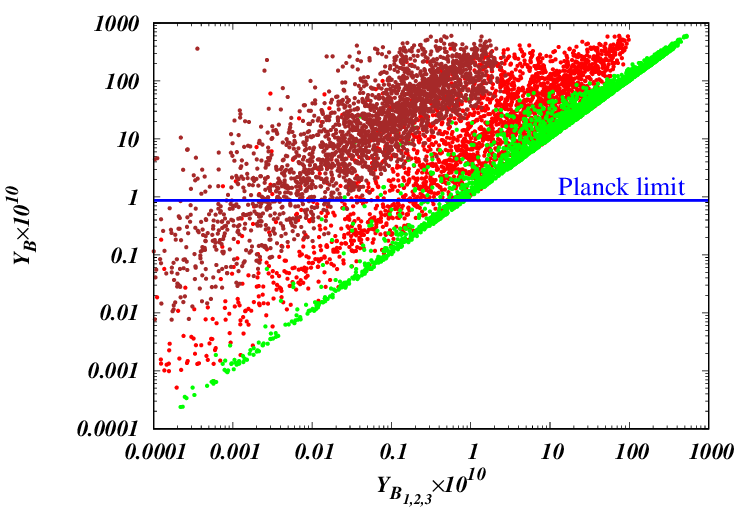} \includegraphics[width=.43%
\textwidth]{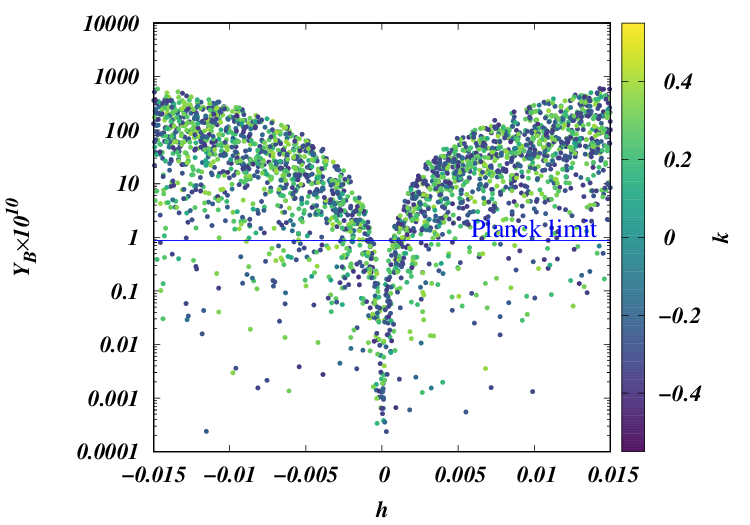}
\end{center}
\par
\vspace{-2em}
\caption{ Left panel: the total baryon asymmetry $Y_{B}$ as a function of $%
Y_{B_{1}}$ (green points), $Y_{B_{2}}$ (red points) and $Y_{B_{3}}$ (brown
points). Right panel: correlation between $Y_{B}$\ and the parameter $h$,
the palette corresponds to the deviation parameter \textrm{k}. The
horizontal blue band corresponds to the Planck bound.}
\label{12}
\end{figure}

\begin{figure}[h]
\begin{center}
\includegraphics[width=.43\textwidth]{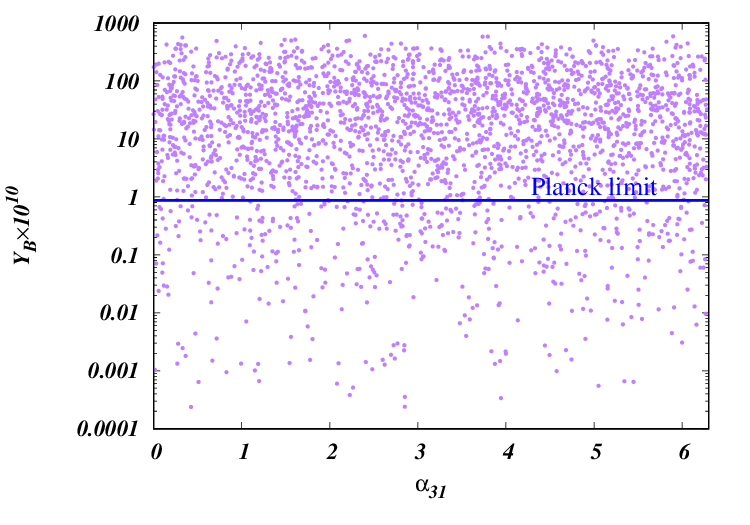} \includegraphics[width=.43%
\textwidth]{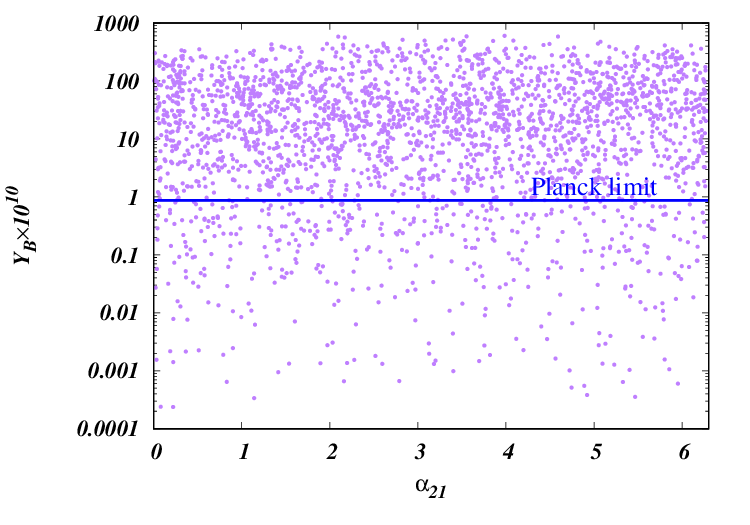} \includegraphics[width=.43\textwidth]{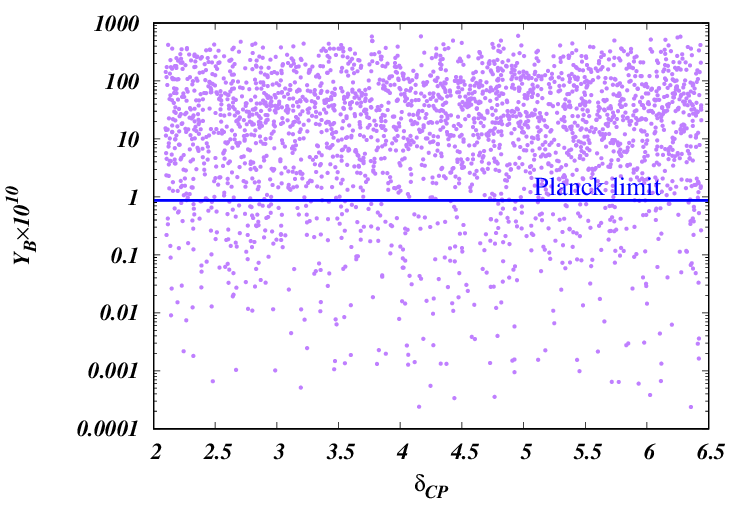} %
\includegraphics[width=.43\textwidth]{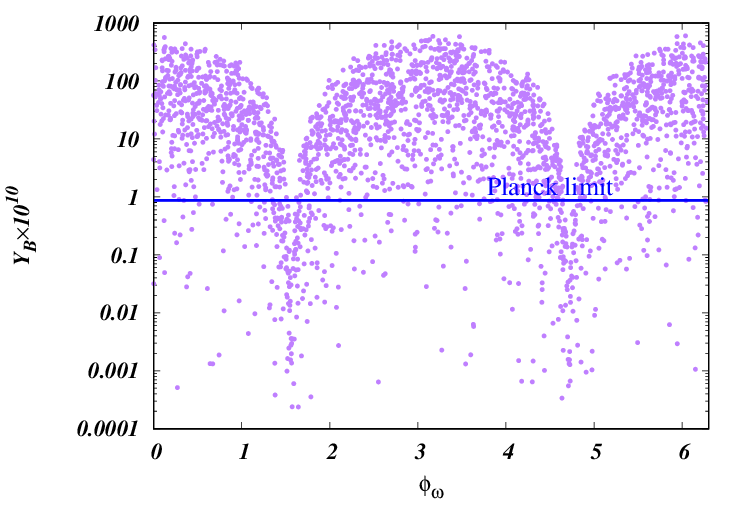}
\end{center}
\par
\vspace{-2em}
\caption{{}The baryon asymmetry $Y_{B}$\ as a function of low energy and
high energy phases. The top-left and -right panels show respectively $Y_{B}$
versus Majorana phases\ $\protect\alpha _{31}$ and $\protect\alpha _{21}$.
The bottom-left panel shows $Y_{B}$ versus Dirac phase $\protect\delta _{CP}$%
. The bottom-right panel shows\ $Y_{B}$ versus the high energy phase $%
\protect\phi _{\protect\omega }$. The total horizontal blue band correspond
to the Planck bound.}
\label{133}
\end{figure}

To estimate the total baryon asymmetry $Y_{B}$\ in our model, we use the
range of the parameters $\theta $, $\gamma $, $m_{i}$, $\alpha _{31}$\ and $%
\alpha _{21}$ allowed by neutrino experiments in the NH case. Furthermore,
since the remaining parameters $h$\ and the phase $\phi _{\omega }$ are not
affected by neutrino oscillation data, we explore their entire ranges of $%
\left[ -0.1\rightarrow 0.1\right] $ and $\left[ 0\rightarrow 2\pi \right] $
respectively. To visualize the correlation between the total baryon
asymmetry $Y_{B}$ and the parameter $h$, we present a plot in the right
panel of Figure (\ref{12}), where the color palette represents the deviation
parameter \textrm{k}. We find that the current observation of the baryon
asymmetry $Y_{B}$ leads to a narrow constraint on the parameter $h$, which
falls within the interval of $\left[ -0.015\rightarrow -0.0008\right] \cup %
\left[ 0.0008\rightarrow 0.015\right] $. Meanwhile, the region for the
deviation parameter \textrm{k} remains unchanged, as discussed in the
previous section.
To analyze the individual contributions $Y_{B_{i}}$ of the \textit{i}th RH
neutrino to the total baryon asymmetry $Y_{B}$, we present a plot in the
left panel of Figure (\ref{12}) showing the dependence of $Y_{B}$ on the
individual parts $Y_{B_{i}}$. We observe that $Y_{B_{1}}$, originating from
the decay of the first RH neutrino, dominates the baryon asymmetry $Y_{B}$.
Meanwhile, the contribution $Y_{B_{3}}$, arising from the lightest RH
neutrino decay, is consistently smaller than both $Y_{B_{1}}$ and $Y_{B_{2}}$%
. For instance, when considering the value $Y_{B}\simeq 8.7007\times
10^{-11} $, which satisfies the experimental observations, it corresponds to
the three contributions $Y_{B_{1}}\simeq 6.71601\times
10^{-11},Y_{B_{2}}\simeq 1.97635\times 10^{-11},Y_{B_{3}}\simeq
0.00832\times 10^{-11}$.\newline
The baryon asymmetry $Y_{B}$ depends on two types of phases; the low energy $%
CP$ phases $\delta _{CP}$, $\alpha _{31}$\ and $\alpha _{21}$ contained in
the lepton mixing matrix and the high energy $CP$ phase $\phi _{\omega }$
originated from the complex coupling constant $\lambda _{9}$ in the Dirac
mass matrix. Therefore, in Figure (\ref{133}), we plot the total baryon
asymmetry $Y_{B}$\ against the four phases $\delta _{CP}$, $\alpha _{31}$%
\textrm{, }$\alpha _{21}$\ and $\phi _{\omega }$. We observe that the hole
inserted interval of the low energy phases phases $\delta _{CP}$, $\alpha
_{31}$ and $\alpha _{21}$\ including its $CP$ conserving values is
consistent with the observations. On the other hand, as shown in the bottom
right panel of Figure (\ref{133}), the region of the high energy phase $\phi
_{\omega }$ varies within the interval $\left[ 0,2\pi \right] $, with the $%
CP $ conserving values $\phi _{\omega }=\frac{\pi }{2},\frac{3\pi }{2}$, and
the surrounding regions being excluded. Therefore, the\ contribution of the
high energy $CP$ phase $\phi _{\omega }$ in $Y_{B}$ plays a subdominant role
in the production of baryon asymmetry compatible with the observations

\begin{figure}[h]
\begin{center}
\includegraphics[width=.45\textwidth]{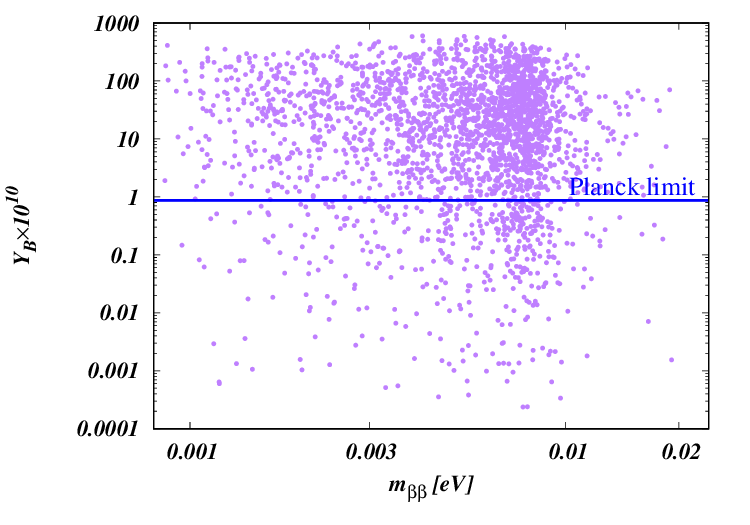}\includegraphics[width=.45%
\textwidth]{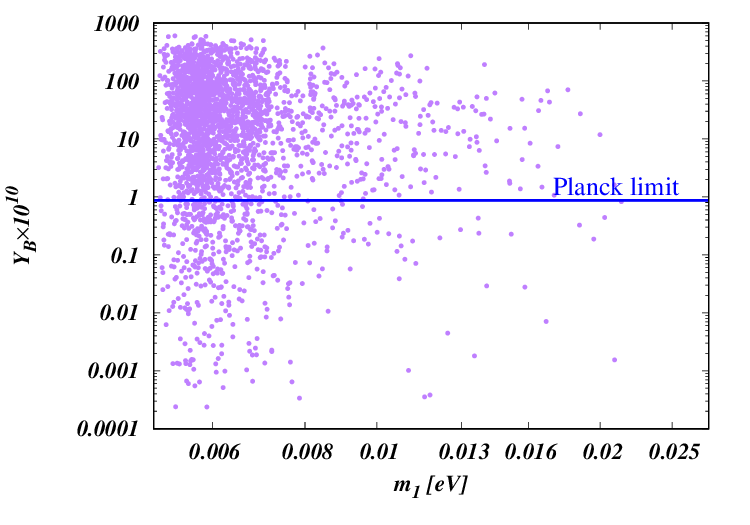} \includegraphics[width=.45\textwidth]{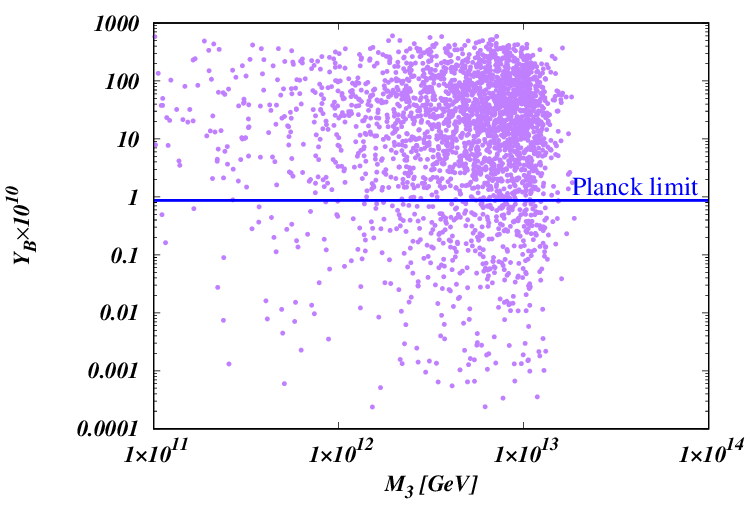}
\end{center}
\par
\vspace{-2em}
\caption{Scatter plot of the total baryon asymmetry $Y_{B}$\ against the
effective Majorana mass $m_{\protect\beta \protect\beta }$ (top-left panel),
the lightest neutrino mass $m_{1}$ (top-right panel)\ and the lightest RH
neutrino mass $M_{3}$ (bottom panel). The total horizontal blue band
correspond to the Planck bound.}
\label{14}
\end{figure}

Given that the total baryon asymmetry $Y_{B}$\ and the effective Majorana
mass $m_{\beta \beta }$ are both sensitive to the Majorana phases, we show
in the top left panel of Figure (\ref{14}) the correlation between these two
observables. We observe that $m_{\beta \beta }$\ maintains the same interval
as obtained in the previous section and has several values that produce the
correct baryon asymmetry. In addition, we plot in Figure (\ref{14}) the
baryon asymmetry $Y_{B}$ versus the lightest neutrino mass $m_{1}$ (top
right panel) and the lightest RH neutrino mass $M_{3}$ (bottom panel). We
observe that the obtained ranges of $m_{1}$\ and $M_{3}$ are consistent with
the observed baryon asymmetry.

\section{Conclusion}

\label{sec5}

We have presented a new and a predictive model based on $D_{4}\times U(1)$
flavor symmetry. Through an analytical and numerical analysis, we have shown
that the model is able to successfully account for the observed neutrino
masses, mixing angles, and the baryon asymmetry of the universe
simultaneously. The model leads to a diagonal charged lepton and Dirac
neutrino mass matrices, together with a heavy Majorana neutrino mass matrix%
\textbf{.} Using the type I seesaw mechanism, a neutrino mass matrix with
broken $\mu -\tau $ symmetry emerges naturally, giving rise to predictive
features concerning the neutrino mixing angles. Furthermore, using the $%
3\sigma $ experimental regions of the neutrino oscillation parameters, we
constrained the model parameter space where our analysis showed that the
model has imperative predictions for neutrino masses and mixing. Our
findings indicate that our model aligns with the observed data exclusively
in the NH scheme, with the atmospheric angle $\theta _{23}$ lies in the
lower octant. A comprehensive investigation of the neutrino phenomena was
carried out, considering non oscillation methods to make predictions about
the absolute neutrino mass scale. Our study involved generating scatter
plots to make several predictions, with the significant finding being that
the effective Majorana neutrino mass, $m_{\beta \beta }$, falls within the
range $[0.567,22.121]$ meV. This region can be tested in upcoming
neutrinoless double beta decay experiments.\newline
On the other hand, to account for the baryon asymmetry, we have investigated
the leptogenesis from the decay of all three RH neutrinos $N_{i}$. Given
that the baryon asymmetry cannot be generated at LO, we have introduced NLO
correction to the Dirac Yukawa matrix involving a new flavon field $\omega $%
. This correction did not have a significant impact on the neutrino sector,
however, it did lead to the emergence of a high-energy phase $\phi _{\omega
} $, which introduced a new source of $CP$ violation. Through a scatter
plot, we have shown that the RH neutrino masses are not highly hierarchical (%
$M_{1}\sim M_{2}\sim 3M_{3}$) and that most of the data points that satisfy
the observed neutrino oscillation fall above the $M_{i}\gtrsim 10^{12}$
bound. Therefore, we have estimated the baryon asymmetry parameter $Y_{B}$
in the unflavored approximation. The numerical results of the baryon
asymmetry $Y_{B}$ have been illustrated through several plots, revealing
that $Y_{B_{1}}$, arising from the RH neutrino $N_{1}$, constitutes the most
significant contribution to the total baryon asymmetry $Y_{B}$.
Nevertheless, we emphasized that the contributions of $Y_{B_{2}}$ and $%
Y_{B_{3}}$ should not be neglected, as they play a crucial role in the
calculation of the BAU in agreement with the Planck limit. Finally, by
varying all parameters in their allowed ranges, we have shown that the high
energy phase $\phi _{\omega }$\ varies within the interval $\left[ 0,2\pi %
\right] $ while the $CP$ conserving values $\phi _{\omega }=\frac{\pi }{2},%
\frac{3\pi }{2}$ and the regions around them are excluded. Therefore, the
high energy $CP$ phase $\phi _{\omega }$\ emerges as a new source of $CP$
violation needed to generate BAU in the current model.\appendix

\section{Dihedral $D_{4}$ group}

\label{app1} The dihedral discrete group $D_{4}$ is a finite group that is
generated by two non commuting elements $S$ and $T$ satisfying the relations
$S^{4}=T^{2}=Id$ and $STS=T$. This group has five irreducible
representations; four singlets denoted as $1_{+,+}$, $1_{+,-}$ $1_{-,+}$ and
$1_{-,-}$, and one doublet $2_{0,0}$ where the sum of their squared
dimensions equal to the order of the $D_{4}$ group through the formula $%
1_{+,+}^{2}+1_{+,-}^{2}+1_{-,+}^{2}+1_{-,-}^{2}+2_{0,0}=8$. We should
mention that the indices of irreducible representations refer to their
characters under the two generators $S$ and $T$ as in the following Table
\cite{C1}%
\begin{equation}
\begin{tabular}{|l|l|l|l|l|l|}
\hline\hline
$\chi _{R_{i}}$ & $\chi _{2_{0,0}}$ & $\chi _{1_{+,+}}$ & $\chi _{1_{+,-}}$
& $\chi _{1_{-,+}}$ & $\chi _{1_{-,-}}$ \\ \hline\hline
$T$ & $\ 0$ & $+1$ & $+1$ & $-1$ & $-1$ \\ \hline
$S$ & $\ 0$ & $+1$ & $-1$ & $+1$ & $-1$ \\ \hline\hline
\end{tabular}%
\end{equation}%
Concerning the tensor products among the irreducible representations of $%
D_{4}$. The tensor product between two $D_{4}$ doublets $(x_{1},x_{2})^{T}$
and $(y_{1},y_{2})^{T}$ is decomposed into a sum of the four singlet
representations of $D_{4}$ as%
\begin{eqnarray}
\left(
\begin{array}{c}
x_{1} \\
x_{2}%
\end{array}%
\right) _{2_{0,0}}\otimes \left(
\begin{array}{c}
y_{1} \\
y_{2}%
\end{array}%
\right) _{2_{0,0}} &=&\left( x_{1}y_{2}+x_{2}y_{1}\right) _{1_{+,+}}\oplus
\left( x_{1}y_{1}+x_{2}y_{2}\right) _{1_{+,-}}\oplus \left(
x_{1}y_{1}-x_{2}y_{2}\right) _{1_{-,+}}  \notag \\
&&\oplus \left( x_{1}y_{2}-x_{2}y_{1}\right) _{1_{-,-}}  \label{c1}
\end{eqnarray}%
whereas the tensor products among the singlet representations can be
expressed as%
\begin{equation}
1_{i,j}\otimes 1_{k,l}=1_{ik,jl}\qquad \text{with}\qquad i,j,k,l=\pm
\label{c2}
\end{equation}

\section{Vacuum alignments for flavon doublets}

\label{app2} In this appendix we discuss the minimization of the scalar
superpotential leading to the alignments of the flavon doublet VEVs. These
VEV alignments are necessary to achieve the desired structures of the
charged lepton and neutrino mass matrices in our model.\ As discussed in
\cite{C2}, the choice of the flavon field directions can be realized by
introducing extra scalar fields with vanishing VEVs called "driving fields"
. This approach employs the continuous $U(1)_{R}$\ symmetry under which the
Higgs and flavon fields have zero charge, the matter fields carry charge $+1$
while the additional driving fields carry charge $+2$. Accordingly, all
terms in the superpotential either contain two matter superfields or one
driving field. Therefore, it is clear that the superpotentials $\mathcal{W}%
_{l}$ and $\mathcal{W}_{\nu }$ in Eqs. (\ref{Wl}) and (\ref{Wn}) are also
invariant under the $U(1)_{R}$ symmetry. Following this approach, we
introduce two driving fields $\digamma ^{0}$\ and $\Omega ^{0}$ which
transform under $(D_{4},U(1))$ as\textrm{\ }%
\begin{equation}
\digamma ^{0}\sim (1_{-,+},-1)\text{\quad and\quad }\Omega ^{0}\sim
(1_{-,-},-4)
\end{equation}%
These scalar fields are assumed to have vanishing VEVs while they are
responsible for aligning the flavon doublets contributing to the charged
lepton and neutrino sectors. Under these assumptions, the renormalizable
superpotential involving the driving fields necessary for aligning the
flavon doublets is given by%
\begin{equation}
\mathcal{W}_{s}=y_{1}\digamma ^{0}\chi \psi +y_{2}\Omega ^{0}\sigma \eta
+y_{3}\Omega ^{0}\rho _{2}\rho _{3}+y_{4}\Omega ^{0}\sigma ^{2}+y_{5}\Omega
^{0}\eta ^{2}
\end{equation}%
where $y_{1,2,3}$ are the coupling constants with absolute values of order
one. The flavon doublets are expressed in terms of $D_{4}$ components as $%
\chi =(\chi _{1},\chi _{2})^{T}$, $\psi =(\psi _{1},\psi _{2})^{T}$, $\sigma
=(\sigma _{1},\sigma _{2})^{T}$ and $\eta =(\eta _{1},\eta _{2})^{T}$.
Therefore, using $D_{4}$ tensor product the superpotential $\mathcal{W}_{s}$
is expressed as
\begin{eqnarray}
\mathcal{W}_{s} &=&y_{1}\digamma ^{0}(\chi _{1}\psi _{1}-\chi _{2}\psi
_{2})+y_{2}\Omega ^{0}(\sigma _{1}\eta _{2}-\sigma _{2}\eta
_{1})+y_{3}\Omega ^{0}\rho _{2}\rho _{3}  \notag \\
&&+y_{4}\Omega ^{0}(\sigma _{1}\sigma _{2}-\sigma _{2}\sigma
_{1})+y_{5}\Omega ^{0}(\eta _{1}\eta _{2}-\eta _{2}\eta _{1})
\end{eqnarray}%
In the limit of unbroken supersymmetry, the vacuum of the flavons is aligned
by setting the F-terms of the driving fields $\digamma ^{0}$\ and $\Omega
^{0}$ to zero as
\begin{eqnarray}
\frac{\partial W}{\partial \digamma ^{0}} &=&y_{1}(\chi _{1}\psi _{1}-\chi
_{2}\psi _{2})=0  \label{f1} \\
\frac{\partial W}{\partial \Omega ^{0}} &=&y_{2}(\sigma _{1}\eta _{2}-\sigma
_{2}\eta _{1})+y_{3}\rho _{2}\rho _{3}+y_{4}(\sigma _{1}\sigma _{2}-\sigma
_{2}\sigma _{1})+y_{5}(\eta _{1}\eta _{2}-\eta _{2}\eta _{1})=0  \label{f2}
\end{eqnarray}%
The first equation involves only doublet flavons which contribute to the
charged lepton masses. It allows clearly for three non trivial solutions
given by%
\begin{eqnarray}
(1) &:&\left\langle \chi \right\rangle =(\upsilon _{\chi },\upsilon _{\chi
})^{T}\ \quad ;\quad \left\langle \psi \right\rangle =(\upsilon _{\psi
},\upsilon _{\psi })^{T}  \notag \\
(2) &:&\left\langle \chi \right\rangle =(0,\upsilon _{\chi })^{T}\ \quad
;\quad \left\langle \psi \right\rangle =(\upsilon _{\psi },0)^{T}
\label{vev} \\
(3) &:&\left\langle \chi \right\rangle =(\upsilon _{\chi },0)^{T}\ \quad
;\quad \left\langle \psi \right\rangle =(0,\upsilon _{\psi })^{T}  \notag
\end{eqnarray}%
The first and the second solutions leads to an inconsistent results
concerning the charged lepton masses. In fact, the first configuration of
VEV leads to vanishing mass $m_{\mu }=0$, this outcome arises due to the
specific form of the charged lepton Yukawa matrix obtained in this scenario,
which is given by the following expression%
\begin{equation}
\mathcal{Y}_{L}=\left(
\begin{array}{ccc}
\lambda _{e}\frac{\upsilon _{\phi }}{\Lambda } & 0 & 0 \\
0 & \lambda _{\mu }\frac{\upsilon _{\chi }}{\Lambda } & \lambda _{\mu }\frac{%
\upsilon _{\chi }}{\Lambda } \\
0 & \lambda _{\tau }\frac{\upsilon _{\psi }}{\Lambda } & \lambda _{\tau }%
\frac{\upsilon _{\psi }}{\Lambda }%
\end{array}%
\right) .
\end{equation}%
The second solution in eq. (\ref{vev}) leads to the charged lepton Yukawa
matrix given by%
\begin{equation}
\mathcal{Y}_{L}=\left(
\begin{array}{ccc}
\lambda _{e}\frac{\upsilon _{\phi }}{\Lambda } & 0 & 0 \\
0 & 0 & \lambda _{\mu }\frac{\upsilon _{\chi }}{\Lambda } \\
0 & \lambda _{\tau }\frac{\upsilon _{\psi }}{\Lambda } & 0%
\end{array}%
\right) .
\end{equation}%
this matrix induces the equality between $m_{\mu }=m_{\tau }$. Conversely,
the last VEV configuration which is the one used in our setup leads to
diagonal charged lepton Yukawa matrix; see eq. (\ref{yl}) with three
hierarchical masses $m_{e}<m_{\mu }<m_{\tau }$. On the other side, we find
that the Eq. (\ref{f2}) admits 6 non trivial solutions for the vacuums$\
\left\langle \eta \right\rangle $\ and $\left\langle \sigma \right\rangle $
which lead to different Majorana mass matrices. They are listed as follows

\begin{itemize}
\item $(1):\left\langle \eta \right\rangle =(\upsilon _{\eta },\upsilon
_{\eta })^{T},\quad \left\langle \sigma \right\rangle =(0,\upsilon _{\sigma
})^{T}$ with $\upsilon _{\eta }=\frac{y_{3}\upsilon _{\rho _{2}}\upsilon
_{\rho _{3}}}{y_{2}\upsilon _{\sigma }}$:%
\begin{equation}
m_{M}=\left(
\begin{array}{ccc}
\lambda _{3}\rho _{1} & \lambda _{5}\eta & \lambda _{5}\eta \\
\lambda _{5}\eta & 0 & 2\lambda _{4}\rho _{1} \\
\lambda _{5}\eta & 2\lambda _{4}\rho _{1} & 0%
\end{array}%
\right) +\left(
\begin{array}{ccc}
0 & 0 & \lambda _{6}\sigma \\
0 & \lambda _{7}\rho _{2}-\lambda _{8}\rho _{3} & 0 \\
\lambda _{6}\sigma & 0 & \lambda _{7}\rho _{2}+\lambda _{8}\rho _{3}%
\end{array}%
\right)  \label{s1}
\end{equation}

\item $(2):\left\langle \eta \right\rangle =(0,\upsilon _{\eta })^{T},\quad
\left\langle \sigma \right\rangle =(\upsilon _{\sigma },\upsilon _{\sigma
})^{T}$ with $\upsilon _{\eta }=-\frac{y_{3}\upsilon _{\rho _{2}}\upsilon
_{\rho _{3}}}{y_{2}\upsilon _{\sigma }}$:%
\begin{equation}
m_{M}=\left(
\begin{array}{ccc}
\lambda _{3}\rho _{1} & \lambda _{6}\sigma & \lambda _{6}\sigma \\
\lambda _{6}\sigma & 0 & 2\lambda _{4}\rho _{1} \\
\lambda _{6}\sigma & 2\lambda _{4}\rho _{1} & 0%
\end{array}%
\right) +\left(
\begin{array}{ccc}
0 & 0 & \lambda _{5}\eta \\
0 & \lambda _{7}\rho _{2}-\lambda _{8}\rho _{3} & 0 \\
\lambda _{5}\eta & 0 & \lambda _{7}\rho _{2}+\lambda _{8}\rho _{3}%
\end{array}%
\right)  \label{s2}
\end{equation}

\item $(3):\left\langle \eta \right\rangle =(\upsilon _{\eta },0)^{T},\quad
\left\langle \sigma \right\rangle =(\upsilon _{\sigma },\upsilon _{\sigma
})^{T}$ with $\upsilon _{\eta }=\frac{y_{3}\upsilon _{\rho _{2}}\upsilon
_{\rho _{3}}}{y_{2}\upsilon _{\sigma }}$:%
\begin{equation}
m_{M}=\left(
\begin{array}{ccc}
\lambda _{3}\rho _{1} & \lambda _{6}\sigma & \lambda _{6}\sigma \\
\lambda _{6}\sigma & 0 & 2\lambda _{4}\rho _{1} \\
\lambda _{6}\sigma & 2\lambda _{4}\rho _{1} & 0%
\end{array}%
\right) +\left(
\begin{array}{ccc}
0 & \lambda _{5}\eta & 0 \\
\lambda _{5}\eta & \lambda _{7}\rho _{2}-\lambda _{8}\rho _{3} & 0 \\
0 & 0 & \lambda _{7}\rho _{2}+\lambda _{8}\rho _{3}%
\end{array}%
\right)
\end{equation}

\item $(4):\left\langle \eta \right\rangle =(\upsilon _{\eta },\upsilon
_{\eta })^{T},\quad \left\langle \sigma \right\rangle =(\upsilon _{\sigma
},0)^{T}$ with $\upsilon _{\eta }=-\frac{y_{3}\upsilon _{\rho _{2}}\upsilon
_{\rho _{3}}}{y_{2}\upsilon _{\sigma }}$:%
\begin{equation}
m_{M}=\left(
\begin{array}{ccc}
\lambda _{3}\rho _{1} & \lambda _{5}\eta & \lambda _{5}\eta \\
\lambda _{5}\eta & 0 & 2\lambda _{4}\rho _{1} \\
\lambda _{5}\eta & 2\lambda _{4}\rho _{1} & 0%
\end{array}%
\right) +\left(
\begin{array}{ccc}
0 & \lambda _{6}\sigma & 0 \\
\lambda _{6}\sigma & \lambda _{7}\rho _{2}-\lambda _{8}\rho _{3} & 0 \\
0 & 0 & \lambda _{7}\rho _{2}+\lambda _{8}\rho _{3}%
\end{array}%
\right)
\end{equation}

\item $(5):\left\langle \eta \right\rangle =(\upsilon _{\eta },0)^{T},\quad
\left\langle \sigma \right\rangle =(0,\upsilon _{\sigma })^{T}$ with $%
\upsilon _{\eta }=\frac{y_{3}\upsilon _{\rho _{2}}\upsilon _{\rho _{3}}}{%
y_{2}\upsilon _{\sigma }}$:%
\begin{equation}
m_{M}=\left(
\begin{array}{ccc}
\lambda _{3}\rho _{1} & \lambda _{5}\eta & \lambda _{6}\sigma \\
\lambda _{5}\eta & \lambda _{7}\rho _{2}-\lambda _{8}\rho _{3} & 2\lambda
_{4}\rho _{1} \\
\lambda _{6}\sigma & 2\lambda _{4}\rho _{1} & \lambda _{7}\rho _{2}+\lambda
_{8}\rho _{3}%
\end{array}%
\right)
\end{equation}

\item $(6):\left\langle \eta \right\rangle =(0,\upsilon _{\eta })^{T},\quad
\left\langle \sigma \right\rangle =(\upsilon _{\sigma },0)^{T}$ with $%
\upsilon _{\eta }=-\frac{y_{3}\upsilon _{\rho _{2}}\upsilon _{\rho _{3}}}{%
y_{2}\upsilon _{\sigma }}$:%
\begin{equation}
m_{M}=\left(
\begin{array}{ccc}
\lambda _{3}\rho _{1} & \lambda _{6}\sigma & \lambda _{5}\eta \\
\lambda _{6}\sigma & \lambda _{7}\rho _{2}-\lambda _{8}\rho _{3} & 2\lambda
_{4}\rho _{1} \\
\lambda _{5}\eta & 2\lambda _{4}\rho _{1} & \lambda _{7}\rho _{2}+\lambda
_{8}\rho _{3}%
\end{array}%
\right)
\end{equation}
\end{itemize}

The Majorana matrices rise from these VEV configurations satisfy the broken $%
\mu -\tau $, leading to consistent predictions on the mixing angles.
Specifically, we have selected the VEV directions as given in $(1)$ along
with the following singlet VEVs%
\begin{equation}
\left\langle \rho _{2}\right\rangle =\upsilon _{\rho _{2}}\quad ,\quad
\left\langle \rho _{3}\right\rangle =\upsilon _{\rho _{3}},\text{\quad }%
\left\langle \eta \right\rangle =(\upsilon _{\eta },\upsilon _{\eta
})^{T},\quad \left\langle \sigma \right\rangle =(0,\upsilon _{\sigma })^{T}
\end{equation}%
These VEVs constitute a stable solution for the second equation in Eq. (\ref%
{f2}). Furthermore, the VEV of the flavon $\eta $ is related to the
remaining flavon VEVs by the relation%
\begin{equation}
\upsilon _{\eta }=\frac{y_{3}\upsilon _{\rho _{2}}\upsilon _{\rho _{3}}}{%
y_{2}\upsilon _{\sigma }}  \label{re}
\end{equation}%
This particular VEV configuration is well-suited for implementing the
trimaximal mixing scheme $TM_{\mathrm{2}}$ in the neutrino sector by
introducing a perturbation matrix $\delta m_{1}$, as shown in Eq. (\ref{per}%
). On the other hand, based on the obtained ranges of the free parameters in
Eq. (\ref{par}), we conclude that all flavon fields introduced in the
neutrino sector have the same order of magnitude. In that regard, the
relation (\ref{re}) emerged from the minimization condition for the field $%
\Omega ^{0}$ implies that the flavon fields are similar to each other in
magnitude which is in agreement with our results. On the other hand, there
is no correlation between the flavon VEVs $\upsilon _{\chi }$ and $\upsilon
_{\psi }$ as indicated by Eq. (\ref{f1}). This is reasonable since they
respectively affect the second and third generations of charged leptons,
which have a hierarchical structure.

\section{Implication of NLO correction $\protect\delta W_{D}$}

\label{app3} The hierarchy of the three lightest neutrino masses $m_{i}$ is
determined in our model by the heavy Majorana masses $M_{i}$ as $\left\vert
m_{i}\right\vert =\frac{\left( \lambda _{1}\upsilon _{u}\right) ^{2}}{M_{i}}$%
. In our study of leptogenesis, we have considered a NLO correction term $%
\delta W_{D}=\frac{\lambda _{9}}{\Lambda }N_{3,2}^{c}L_{\mu ,\tau
}H_{u}\omega $ which involves the new flavon field $\omega $. Thus, the
resulting neutrino mass matrix can be expressed as $m_{\nu }^{\prime
}=m_{\nu }+\delta m_{\nu }$\ where $m_{\nu }$ is the neutrino mass matrix at
leading order while $\delta m_{\nu }$ is the correction given by%
\begin{eqnarray}
\delta m_{\nu } &=&\frac{\upsilon _{u}^{2}h\lambda _{1}}{H\Lambda }%
\allowbreak \left(
\begin{array}{ccc}
0 & -\left( b^{2}-bk+ab-k^{2}\right) & -\left( a+b\right) \left( b+k\right)
\\
-\left( b^{2}-bk+ab-k^{2}\right) & \left( 2a^{2}-2b^{2}+2ab-2bk\right) &
\left( 2b^{2}+k^{2}-ak+2bk\right) \\
-\left( a+b\right) \left( b+k\right) & \left( 2b^{2}+k^{2}-ak+2bk\right) &
\left( 2a^{2}-2b^{2}+2ab-2bk\right)%
\end{array}%
\right)  \notag \\
&&+\frac{\upsilon _{u}^{2}h^{2}}{H\Lambda }\left(
\begin{array}{ccc}
0 & 0 & 0 \\
0 & \left( b^{2}-ak\right) & \left( a^{2}-b^{2}+ab-bk\right) \\
0 & \left( a^{2}-b^{2}+ab-bk\right) & \left( b^{2}+k^{2}+2bk\right)%
\end{array}%
\right)
\end{eqnarray}%
The neutrino masses taking into account the correction $\delta m_{\nu }$ are
given approximately as%
\begin{eqnarray}
m_{1}^{\prime } &\simeq &\frac{\left( \lambda _{1}\upsilon _{u}\right) ^{2}}{%
M_{1}}=m_{1}  \notag \\
m_{2}^{\prime } &\simeq &\frac{\upsilon _{u}^{2}(\lambda _{1}+h)^{2}}{M_{2}}%
=m_{2}+2\frac{\upsilon _{u}^{2}\lambda _{1}h}{M_{2}}+\frac{\upsilon
_{u}^{2}h^{2}}{M_{2}} \\
m_{3}^{\prime } &\simeq &\frac{\upsilon _{u}^{2}(\lambda _{1}-h)^{2}}{M_{3}}%
=m_{3}-2\frac{\upsilon _{u}^{2}\lambda _{1}h}{M_{3}}+\frac{\upsilon
_{u}^{2}h^{2}}{M_{3}}  \notag
\end{eqnarray}%
Given that the parameter $h$ has an origin related to an NLO correction
term, it is expected to be small. In fact, our analysis reveals that to
generate the observed baryon asymmetry, the parameter $h$ falls within the
interval $\left\vert h\right\vert \in \left[ 0.008\rightarrow 0.015\right] $%
. As a result, the contributions $\pm 2\frac{\upsilon _{u}^{2}\lambda _{1}h}{%
M_{i}}+\frac{\upsilon _{u}^{2}h^{2}}{M_{i}}$ are negligible compared to the
LO contributions that are primarily responsible for the neutrino masses.
Accordingly, the correction $\delta m_{\nu }$\ will not provide any
significant impact on the mixing angles $\theta _{ij}$ which are mainly
derived from Majorana matrix within TM$_{2}$.

\end{document}